\documentclass{llncs}
\usepackage{xspace,comment}
\usepackage{amssymb,amscd,tikz}
\usepackage{amsfonts,stmaryrd}
\usepackage{amsmath}
\usepackage[utf8]{inputenc}
\usepackage{graphicx}
\usepackage{pifont}
\usepackage{tabls,subfig}
\usepackage{tikz}
\usepackage{enumerate}
\newcommand{\z}{\ensuremath{\mathbb{Z}}\xspace}

\newcommand{\n}{\ensuremath{\mathbb{N}}\xspace}

\newcommand{\ie}{\emph{i.e.}\@\xspace}

\newcommand{\Zgrid}{{\mathbb Z}^2}

\newcommand{\sv}{S_{\vec v}}

\newcommand{\re}{\mathbb{R}}
\newcommand{\q}{\mathbb{Q}}

\newcommand{\az}{\ensuremath{A^{\mathbb{Z}}}\xspace}
\newcommand{\bz}{\ensuremath{B^{\mathbb{Z}}}\xspace}
\newcommand{\azd}{\ensuremath{A^{\mathbb{Z}^D}}\xspace}
\newcommand{\zdu}{\ensuremath{\mathbb{Z}^2}\xspace}
\newcommand{\azdu}{\ensuremath{A^{\zdu}}\xspace}

\newcommand{\set}[1]{\left\{#1\right\}}
\newcommand{\sh}[1]{\mathbf{[}#1\mathbf{]}} 

\newcommand{\lzs}{L_0^*}
\newcommand{\para}[1]{(#1)}
\newcommand{\nn}{\vec\nu}
\newcommand{\gv}{\vec\gamma}
\newcommand{\T}{\ensuremath{\mathcal{T}}}
\newcommand{\Tr}{\ensuremath{\mathcal{T}}}

\newcommand{\kk}{\vec k}
\renewcommand{\ll}{\vec \lambda}
\newcommand{\mm}{\vec \mu}

\newcommand{\xx}{\vec x}
\newcommand{\yy}{\vec y}

\newcommand{\vv}{\vec v}
\newcommand{\dd}{\vec d}

\newcommand{\Mr}{\ensuremath{\mathcal{M}_r}\xspace}

\newcommand{\wrt}{w.r.t.\@\xspace}

\newcommand{\etc}{\emph{etc.}\@\xspace}
\newcommand{\ignore}[1]{}

\colorlet{a}{white!30!black}
\colorlet{b}{white!45!black}
\colorlet{c}{white!55!black}
\colorlet{d}{white!70!black}
\colorlet{w}{white}
\tikzstyle{help lines}+=[color=black!90]

\begin{document}
\pagestyle{empty}  
\mainmatter
\title{2D cellular automata:\\ dynamics and undecidability
\ignore{
\thanks{This work has been supported by
the PRIN/MIUR project
``Formal Languages and Automata: Mathematical and Applicative Aspects''.} }}
\author{Alberto Dennunzio\inst{1}
\and Enrico Formenti\inst{2}\thanks{Corresponding author.}
\and Michael Weiss\inst{2}}
\institute{
         Universit\`a degli Studi di Milano--Bicocca\\
        Dipartimento di Informatica, Sistemistica e Comunicazione,\\
        Viale Sarca 336, 20126 Milano (Italy)\\
\email{dennunzio@disco.unimib.it}\quad\email{michael.weiss@cui.unige.ch}
        \and
Universit\'e de Nice-Sophia Antipolis,
Laboratoire I3S,\\
2000 Route des Colles, 06903 Sophia Antipolis (France).\\
\email{enrico.formenti@unice.fr}
          }
\maketitle

\begin{abstract}
In this paper we introduce the notion of quasi-expansivity for 2D
CA and we show that it shares many properties with expansivity (that holds only for 1D
CA). Similarly, we introduce the notions of quasi-sensitivity and prove that the classical dichotomy theorem holds in
this new setting. Moreover, we show a tight relation between 
closingness and openness for 2D CA. Finally, the undecidability of
closingness property for 2D CA is proved. 
\end{abstract}

\section{Introduction}
Cellular automata (CA) are a widely used formal model for complex systems with applications in many different fields ranging from
physics to biology, computer science, mathematics, \etc.
Although applications mainly concern two or higher dimensional
CA, the study of the dynamical behavior has been mostly carried on 
in dimension $1$. Only few results are known for dimension $2$, and practically speaking,  a systematic study of 2D CA dynamics has just started (see for example~\cite{theyssier08,dennunzio08}).
This paper contributes the following main results:
\begin{itemize}
\item properties characterizing quasi-expansive 2D CA;
\item topological entropy of quasi-expansive 2D CA is infinite;
\item a dichotomy for quasi-sensitivity.
\item a tight relation between closingness and openness;
\item undecidability of closingness for 2D CA;
\end{itemize}

It is well-known that there is no positively expansive 2D CA~\cite{shereshevsky93}.\ignore{
 This result can be restated in terms of defects propagation. Indeed, in dimension 1, given a pair of distinct configurations $x,y$, call defects differences between $x$ and $y$.
Any expansive CA create and propagates defects in both directions (left and right) at each time step. In this way, at some later time $t$, $F^t(x)$ and $F^t(y)$ will differ in a cell near the
origin (how much near is determined by the expansivity constant).
The same behavior is not possible in dimension 2. The point is that
for any precision $\epsilon>0$, the exists a pair of configurations 
for which defects propagate in the 2D space avoiding the 
square $[-n,n]^2$ centered in the origin ($n$ is such that
$n$ is the least integer such that $2^{-n}<\epsilon$).
However, computer experiments show that there are many 2D CA which are able to create new defects and propagate them at each 
time step although this phenomenon is not enough to provide expansive behavior. In some sense these CA ``seem'' expansive and the fact that they are not appears to be more an artifact of the Cantor metric than an intrinsic property of the automaton.}
However, the absence of positively expansive 2D CA seems, at a certain extent,
more an artifact of Cantor metric than an intrinsic property of CA.
In this paper we introduce a new notion, namely \emph{quasi-expansivity}, to capture this intuition. We prove that quasi-expansivity shares with positive expansivity several properties (Theorems~\ref{th:expclosing},~\ref{th:expmixing} and Proposition~\ref{prop:bipexp}) and it seems to us the good notion for studying ``this kind'' of dynamics in dimension 2 or higher. 

By a result in~\cite{theyssier08}, the classical dichotomy between sensitive and almost equicontinuous CA is no more true in dimension 2 or higher. In this paper, we prove that the dichotomy theorem still holds (Proposition~\ref{prop:quasisens}) if
the notion of sensitivity is suitably changed. 

In~\cite{dennunzio08}, the notion of closingness has been generalized to 2D and higher. Theorem~\ref{th:2closingopen} states that
bi-closing 2D CA are open. This result has many interesting consequences over the dynamical behavior. For example, quasi-expansive 2D CA turn out to be open 
(Corollary~\ref{cor:quasiexp-dpo-surj-op}). As in~\cite{dennunzio08}, most of these results have been obtained
using the slicing construction, confirming it as a powerful tool for the analysis of 2D CA dynamics. We stress that, even if the constructions are of help for
proving 1D--like results, most of the proofs differ significantly from their 1D
counterparts.
In Section~\ref{sec:closund}, we prove that closingness (and some other related to it) is undecidable in the 2D case (Theorem~\ref{th:closingundec}). Remark that this results corrects an error
made in~\cite[Prop. $2$]{dennunzio08} due to a wrong use of the property characterizing closing CA (\cite[Prop. $1$]{dennunzio08}). Recalling that closingness is decidable in
dimension 1 (see~\cite{kurka04}), we have just added one more
item to the slowly growing collection of dimension sensitive properties
(see~\cite{kari94a,bernardi05} for other examples). 
Moreover, the proof technique used for Theorem~\ref{th:closingundec}
generalizes classical Kari's construction~\cite{kari94a} which uses tiling and plane-filling curves. We believe that this new construction is of some interest in its own.
\smallskip

The paper is structured as follows. Next section recalls basic notions and some known results about CA and discrete dynamical systems.
Section~\ref{sec:slicing} presents the slicing construction. 
Sections~\ref{sec:clo-open} to~\ref{sec:closund} contain the main results.

%
%

\section{Basic notions}
In this section we briefly recall standard definitions about CA as
dynamical systems. For introductory matter see~\cite{kurka04}.
%
%
For all $i,j\in\z$ with $i\leq j$ (resp., $i<j$), let
$[i,j]=\set{i,i+1,\ldots,j}$ (resp., $[i,j)=\set{i,i+1,\ldots,j-1}$). 
Let $\n_+$ be the set of positive
integers. For a vector $\xx\in\zdu$, denote by $|\xx|$ the
infinite norm (in $\re^2$) of $\xx$. Let $r\in\n$. Denote by $\Mr$
the set of all the two-dimensional matrices with values in $A$ and
entry vectors in the square $[-r,r]^2$. For any matrix $N\in\Mr$,
$N(\xx)\in A$ represents the element of the matrix with entry
vector $\xx$.
\paragraph{{\bf 1D CA.}} Let $A$ be a possibly infinite alphabet. A \emph{1D CA
configuration} is a function from $\z$ to $A$. The \emph{1D CA
configuration set} $\az$ is usually equipped with the metric $d$
defined as follows
\[
\forall c,c^\prime\in\az,\;
d(c,c^\prime)=2^{-n},\;\text{where}\;n=\min\set{i\geq 0\,:\,c_i\ne
c^\prime_i \;\text{or}\;c_{-i}\ne c^\prime_{-i}}\enspace.
\]
If $A$ is finite, $\az$ is a compact, totally disconnected and
perfect topological space (\ie it is a Cantor space). For any
pair $i,j\in\z$, with $i\leq j$, and any configuration $c\in\az$
we denote by $c_{[i,j]}$ the word $c_i\cdots c_j\in A^{j-i+1}$.
A \emph{cylinder} of block $u\in A^k$ and position $i\in\z$ is the
set $[u]_i=\{c\in A^{\z}: c_{[i, i+k-1]}=u\}$. Cylinders are
clopen sets \wrt the metric $d$ and they form a basis for the
topology induced by $d$.
A \emph{1D CA} is a structure $\langle 1, A, r, f\rangle$, where
$A$ is the alphabet, $r\in\n$ is the \emph{radius} and $f:
A^{2r+1} \to A$ is the \emph{local rule} of the automaton. The
local rule $f$ induces a \emph{global rule} $F:\az\to\az$ defined
as follows,
\[
\forall c\in \az,\,\forall i\in\z ,\quad F(c)_i= f(c_{i-r},
\ldots, c_{i+r})\enspace.
\]
Note that $F$ is a uniformly continuous map \wrt the metric $d$.
A 1D CA with global rule $F$ is \emph{right} (resp., \emph{left})
\emph{closing} iff $F(c)\neq F(c^\prime)$ for any pair
$c,c^\prime\in\az$ of distinct left (resp., right) asymptotic
configurations, \ie, $c_{(-\infty,n]}=c^\prime_{(-\infty,n]}$
(resp., $c_{[n,\infty)}=c^\prime_{[n,\infty)}$) for some $n\in\z$,
where $a_{(-\infty,n]}$ (resp., $a_{[n, \infty)}$) denotes the
portion of a configuration $a$ inside the infinite integer
interval $(-\infty,n]$ (resp., $[n, \infty)$). A CA is said to be
\emph{closing} if it is either left or right closing. A rule
$f:A^{2r+1} \to A$ is \emph{righmost} (resp., \emph{leftmost})
\emph{permutive} iff $\forall u\in A^{2r}, \forall\beta\in
A,\exists! \alpha\in A$ such that $f(u\alpha)=\beta$ (resp.,
$f(\alpha u)=\beta$).
\paragraph{{\bf 2D CA.}} Let $A$ be a finite alphabet. A \emph{2D CA
configuration} is a function from $\zdu$ to $A$. The \emph{2D CA
configuration set} $\azdu$ is equipped with the following metric
which is denoted for the sake of simplicity by the same symbol of
the 1D case:
\[
  \forall c,c^\prime\in\azdu,\quad d(c,c^\prime)=2^{-k}
  \quad\text{where}\quad k=\min\set{|\xx|:
  \xx\in \zdu, c(\xx)\ne c^\prime(\xx)}\enspace.
\]
The 2D configuration set is a Cantor space.
A \emph{2D CA} is a structure $\langle 2, A, r, f\rangle$, where
$A$ is the alphabet, $r\in\n$ is the \emph{radius} and $f: \Mr \to
A$ is the \emph{local rule} of the automaton. The local rule $f$
induces a \emph{global rule} $F:\azdu\to\azdu$ defined as follows,
\[
\forall c\in \azdu,\,\forall \xx\in\zdu ,\quad F(c)(\xx)=
f\big(M^{\xx}_r(c)\big)\enspace ,
\]
where $M^{\xx}_r(c)\in\Mr$ is the \emph{finite portion} of $c$
with center $\xx\in\zdu$ and radius $r$ defined by $\forall
\kk\in [-r,r]^2$, $M^{\xx}_r(c)(\kk)=c(\xx+\kk)$.
For any $\vv\in\zdu$ the \emph{shift map}
 $\sigma^{\vv}: A^{\zdu} \to A^{\zdu}$ is defined
by $\forall c\in\azdu, \forall \xx\in\zdu$,
$\sigma^{\vv}(c)(\xx)=c(\xx+\vv)$. A function $F:\azdu \to \azdu$
is said to be \emph{shift-commuting} if $\forall \kk\in\zdu$,
$F\circ\sigma^{\kk}=\sigma^{\kk}\circ F$. Note that 2D CA are
exactly the class of all shift-commuting functions which are
(uniformly) continuous with respect to the metric $d$\ignore{(Hedlund's
theorem from~\cite{hedlund69}). A 2D \emph{subshift} $S$ is a
closed subset of the CA configuration space such that for any
$\vv\in\zdu$, $\sigma^{\vv}(S)\subset S$}.
For any fixed vector $\vv$, we denote by $\sv$ the set of all
configurations $c\in\azdu$ such that $\sigma^{\vv}(c)=c$. Remark
that, for any 2D CA global map $F$ and for any $\vv$, the set
$\sv$ is $F$-invariant, \ie, $F(\sv)\subseteq \sv$.
\paragraph{{\bf DTDS.}} A \emph{discrete time dynamical system (DTDS)}
is a pair $\para{X,g}$ where $X$ is a set equipped with a distance
$d$ and $g:X\mapsto X$ is a map which is continuous on $X$ with
respect to the metric $d$. When $X$ is the configuration space of
a (either 1D or 2D) CA equipped with the above introduced metric,
the pair $\para{X,F}$ is a DTDS. From now on, for the sake of
simplicity, we identify a CA with the dynamical system induced by
itself or even with its global rule $F$.
Given a DTDS $\para{X,g}$, an element $c\in X$ is
 an \emph{equicontinuity point} for $g$ if $\forall\varepsilon>0$
there exists $\delta>0$ such that for all $c^\prime\in X$,
$d(c^\prime,c)<\delta$ implies that $\forall
n\in\n,\;d(g^n(c^\prime),g^n(c))<\varepsilon$. For a 1D CA $F$,
the existence of an equicontinuity point is related to the
existence of a special word, called \emph{blocking word}. A word
$u\in A^k$ is $s$-blocking ($s\leq k$) for a CA $F$ if there
exists an offset $j\in [0, k-s]$ such that for any $x,y\in [u]_0$
and any $n\in\n$, $F^n(c)_{[j,j+s-1]}=F^n(c^\prime)_{[j,j+s-1]}$\,. A
word $u\in A^k$ is said to be \emph{blocking} if it is
$s$-blocking for some $s\leq k$. 
A DTDS is said to be
\emph{equicontinuous} if $\forall\varepsilon>0$ there exists
$\delta>0$ such that for all $c,c^\prime\in X$,
$d(c^\prime,c)<\delta$ implies that $\forall
n\in\n,\;d(g^n(c^\prime),g^n(c))<\varepsilon$. \ignore{If $X$ is a compact
set, a DTDS $\para{X,g}$ is equicontinuous iff the set $E$ of all
its equicontinuity points is the whole $X$.} A DTDS is said to be
\emph{almost equicontinuous} if the set $E$ of its equicontinuity points is residual (\ie, $E$ contains a countable intersection of dense open subsets).
Recall that a DTDS $\para{X,g}$ is \emph{sensitive to the initial
conditions} (or simply \emph{sensitive}) if there exists a
constant $\varepsilon>0$ such that for any $c\in X$ and any
$\delta>0$ there is an element $c^\prime\in X$ such that
$d(c^\prime,c)<\delta$ and $d(g^n(c^\prime),g^n(c))>\varepsilon$
for some $n\in\n$. In~\cite{kurka97}, K\r{u}rka proved that a 1D CA on a finite alphabet is almost equicontinuous iff it is non-sensitive iff it admits a $r$-blocking word.
A DTDS $\para{X,g}$ is \emph{positively
expansive} if there exists a constant $\varepsilon>0$ such that
for any pair of distinct elements $c,c^\prime$ we have
$d(g^n(c^\prime),g^n(c))\geq\varepsilon$ for some $n\in\n$.

Given a DTDS $\para{X,g}$, a point $c\in X$ is \emph{periodic} for
$g$ if there exists an integer $p>0$ such that $g^p(c)=c$. If the
set of all periodic points of $g$ is dense in $X$, we say that the
DTDS has the \emph{denseness of periodic orbits (DPO)}.
%
%
\ignore{
Recall that a DTDS $\para{X,g}$ is \emph{(topologically)
transitive} if for any pair of non-empty open sets
$O_1,O_2\subseteq X$ there exists an integer $n\in\n$ such that
$g^n(O_1)\cap O_2\ne\emptyset$.}
Recall that a DTDS $\para{X,g}$ is \emph{(topologically) mixing} if 
for any pair of non-empty open sets $U,V\subseteq X$ there 
exists an integer $n\in\n$ such that for any $t\geq n$ we have 
$g^t(U)\cap V\ne\emptyset$. Recall that a DTDS $\para{X,g}$ is \emph{(topologically) strongly transitive} if 
for any non-empty open set $U$ it holds that $\bigcup_{n\in\n} g^n(U)=X$.
A DTDS $\para{X,g}$ is \emph{open}
(resp., \emph{surjective}) iff  $g$ is open (resp., $g$ is
surjective).
Recall that two DTDS $\para{X,g}$ and $\para{X^\prime,g^{\prime}}$
are \emph{isomorphic} (resp., \emph{topologically conjugated}) if
there exists a bijection (resp., homeomorphism) $\phi:X\mapsto
X^\prime$ such that
$g^\prime\circ \phi=\phi\circ g$. $\para{X^\prime,g^{\prime}}$ is
a \emph{factor} of $\para{X,g}$ if  there exists a continuous and
surjective map $\phi:X\to X^\prime$ such that $g^\prime\circ
\phi=\phi\circ g$.
Remark that in that case, $\para{X^\prime,g^{\prime}}$ inherits
from $\para{X,g}$ some properties such as surjectivity,
mixing, and DPO.

\section{A powerful tool: the slicing construction}
\label{sec:slicing}

We review two powerful constructions for CA in
dimension greater than 1. The idea inspiring these constructions
appeared in the context of additive CA in~\cite{margara99} and it
was formalized in~\cite{CDM04}.  We
generalize it to arbitrary 2D CA. Moreover, we further refine it
so that slices are translation invariant along some fixed
direction. This confers finiteness to the set of states of the
sliced CA allowing to lift even more properties.
\smallskip

The constructions are given with respect to any direction for 2D
CA, improving the ones introduced in~\cite{dennunzio08}. The
generalization to higher dimensions is straightforward.
%
\smallskip

Fix a vector $\nn\in\zdu$ and let $\dd\in\zdu$ be a normalized
integer vector (\ie a vector with co-prime coordinates)
perpendicular to $\nn$. Consider the line $L_0$ generated by the
vector $\dd$ and the set $\lzs=L_0\cap \zdu$ containing vectors of
form $\xx= t\vec d$ where $t\in\z$. Denote by
$\varphi:\lzs\mapsto\z$ the isomorphism associating any
$\xx\in\lzs$ with the integer $\varphi(\xx)= t$. Consider now the
family $\mathcal{L}$ constituted by all the lines parallel to
$L_0$ containing at least a point of integer coordinates.
It is clear that $\mathcal{L}$ is in a one-to-one correspondence
with $\z$. Let $l_a$ be the axis given by a direction ${\vec e}_a$
which is not contained in $L_0$. We enumerate the lines according
to their intersection with the axis $l_a$. Formally, for any pair
of lines $L_i, L_j$, it holds that $i<j$ iff $p_i<p_j$
($p_i,p_j\in\q$), where $p_i{\vec e_a}$ and $p_j{\vec e_a}$
 are the intersection points between the two lines
and the axis $l_a$, respectively. Equivalently, $L_i$ is the line
expressed in parametric form by $\xx=p_i\vec e_a+t\dd$
($\xx\in\re^2, t\in\re$) and $p_i=ip_1$, where $p_1=\min\set{p_i,
p_i>0}$. Remark that $\forall i,j\in\z$, if $\xx\in L_i$ and
$\yy\in L_j$, then $\xx+\yy\in L_{i+j}$. Let $\yy_1\in \zdu$ be an
arbitrary but fixed vector of $L_1$. For any $i\in\z$, define the
vector $\yy_i=i\yy_1$ which belongs to $L_i\cap \zdu$. Then, each
line $L_i$ can be expressed in parametric form by
$\xx=\yy_i+t\dd$. Note that, for any $\xx\in \zdu$ there exist
$i,t\in\z$, such that $\xx=\yy_i+ td$.
\begin{figure}[!htb]
  \begin{center}
     \includegraphics[scale=.4]{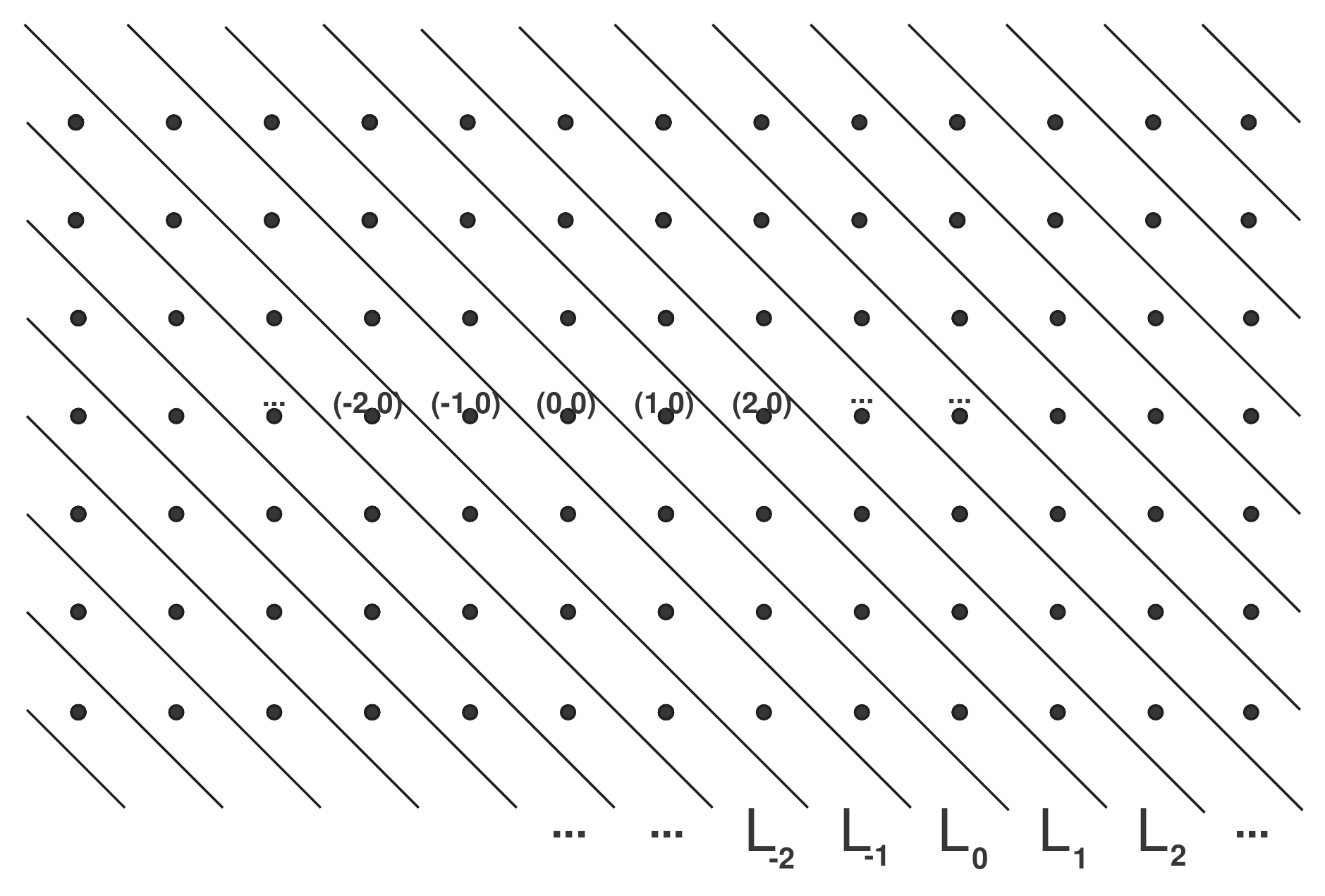}
   \end{center}
   \caption{Slicing of the plane according to the vector $\nn=(1,1)$.}
 \label{fig:slicing-plane}
\end{figure}
Let us summarize the construction. We have a countable collection
${\mathcal{L}}=\{L_i: i\in\z\}$ of lines parallel to $L_0$
inducing a partition of $\zdu$. Indeed, defining
$L^*_i=L_i\cap\zdu$, it holds that $\zdu=\bigcup_{i\in\z} L_i^*$
(see Figure~\ref{fig:slicing-plane}).

\begin{figure}[!htb]
  \begin{center}
     \includegraphics[scale=.4]{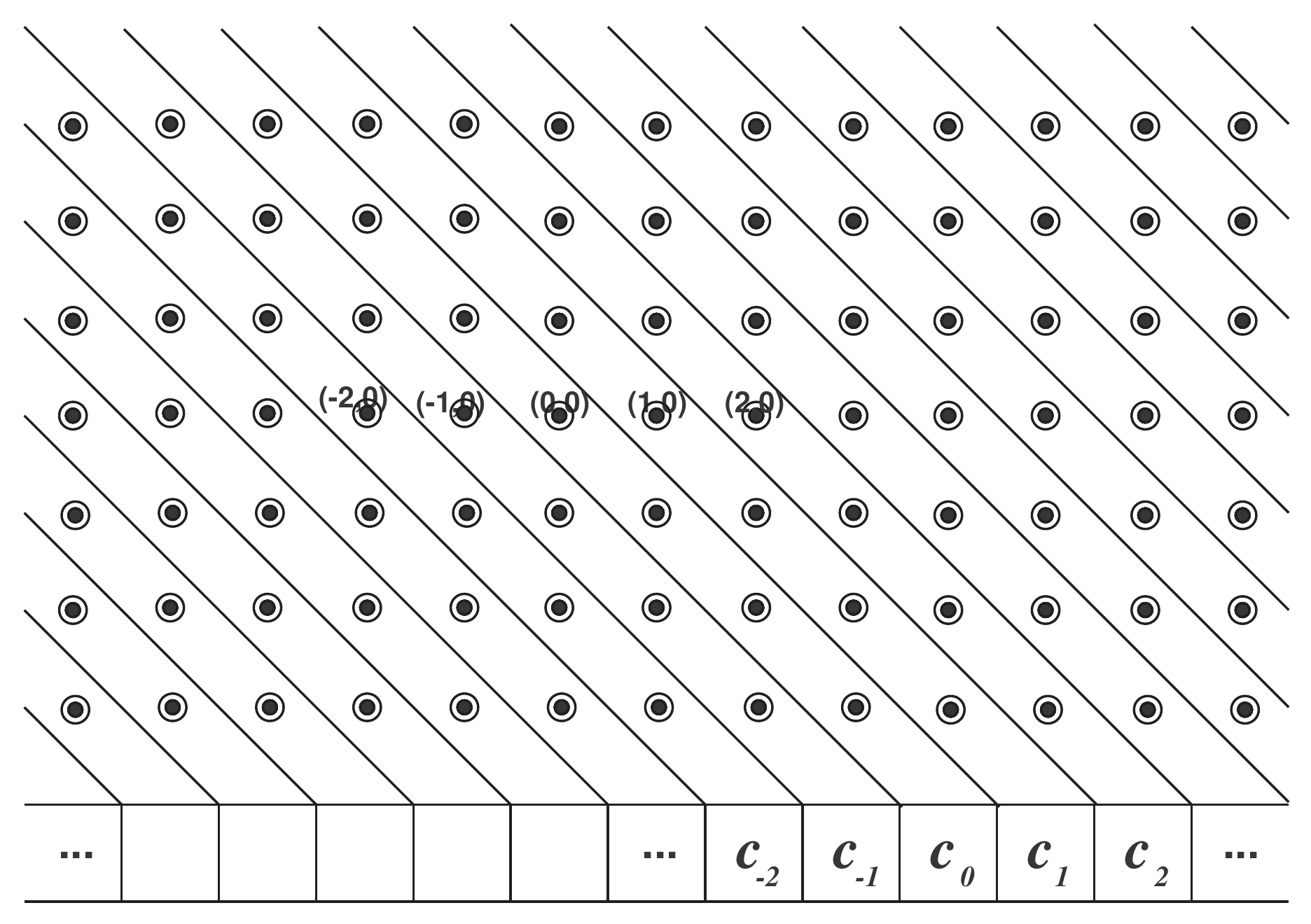}
   \end{center}
   \caption{Slicing of a 2D configuration $c$ according to the vector $\nn=(1,1)$.
   The components of $c$ viewed as a 1D configuration
   are not from the same alphabet.}
  \label{fig:slicing-conf}
\end{figure}

Once the plane has been sliced, any configuration $c\in \azdu$ can
be viewed as a mapping $c:\bigcup_{i\in\z}L_i^*\mapsto\z$. For
every $i\in\z$, the \emph{slice} $c_i$ of the configuration $c$
over the line $L_i$ is the mapping $c_i:L^*_i\to A$. In other
terms, $c_i$ is the restriction of $c$ to the set
$L^*_i\subset\zdu$. In this way, a configuration $c\in\azdu$ can
be expressed as the bi-infinite one-dimensional sequence $\prec
c\succ=(\ldots, c_{-2}, c_{-1}, c_{0}, c_{1}, c_2,\ldots)$ of its
slices $c_i\in A^{L^*_i}$ where the $i$-th component of the
sequence $\prec c\succ$ is $\prec c\succ_i=c_i$ (see
Figure~\ref{fig:slicing-conf}). Let us stress that each slice
$c_i$ is defined only over the set $L^*_i$. Moreover, since
$\forall \xx\in\zdu,\,\exists ! i\in\z: \xx\in L^*_i$, for any
configuration $c$ and any vector $\xx$ we write $c(\xx)=c_i(\xx)$.

The identification of any configuration $c\in\azdu$ with the
corresponding bi-infinite sequence of slices $c \equiv \prec
c\succ=(\ldots, c_{-2}, c_{-1}, c_{0}, c_{1}, c_2, \ldots)$,
allows the introduction of a new one-dimensional bi-infinite CA
over the alphabet $\az$ expressed by a global transition mapping
$F^*:(\az)^{\z}\mapsto(\az)^{\z}$ which associates any
configuration $a:\z\mapsto\az$ with a new configuration
$F^*(a):\z\to\az$. The local rule $f^*$ of this new CA we are
going to define will take a certain number of configurations of
$\az$ as input and will produce a new configuration of $\az$ as
output.

\begin{figure}[!htb]
  \begin{center}
     \includegraphics[scale=.4]{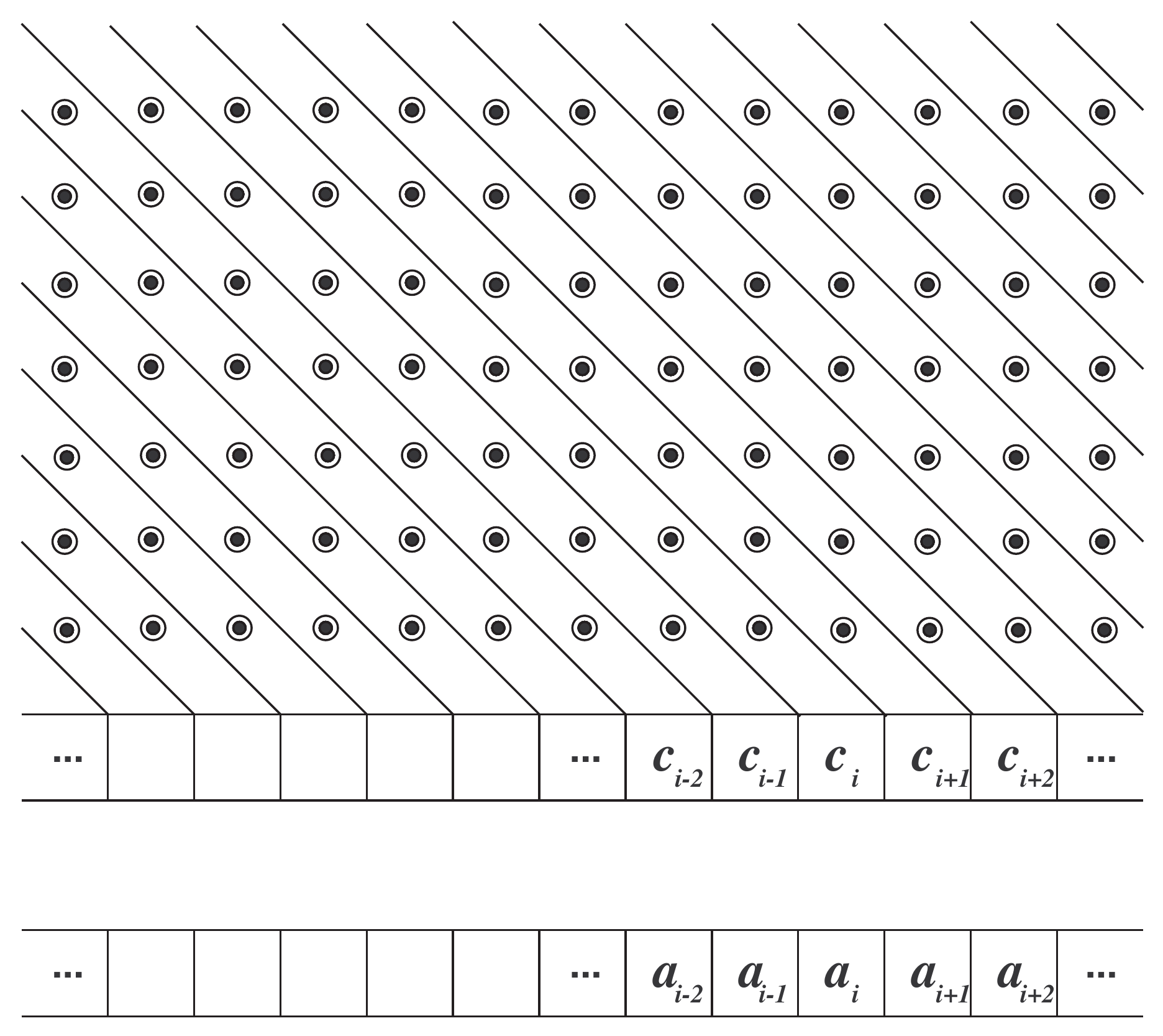}
   \end{center}
   \caption{All the components of the 1D configuration
   $a=\Phi(c)$ are from the same alphabet.}
 \label{fig:same-alph}
\end{figure}

For each $h\in\z$, define the following bijective map $\Tr_{h}:
A^{L^*_{h}}\mapsto A^{L^*_{0}}$ which associates any slice $c_{h}$
over the line $L_{h}$ with the slice $\Tr_{h}(c_{h})$
\[
  \left( c_{h}: L^*_{h}\to A\right)\quad \xrightarrow{\Tr_{h}}
 \quad \left(\Tr_{h}(c_{h}):
 L^*_{0}\to A\right)
\]
defined as $\forall \xx\in L^*_{0},
\Tr_{h}(c_{h})(\xx)=c_{h}(\xx+\yy_h).$ Remark that the map
$\Tr^{-1}_h: A^{L^*_{0}}\to A^{L^*_{h}}$ associates any slice
$c_{0}$ over the line $L_{0}$ with the slice $\Tr^{-1}_{h}(c_{0})$
over the line $L_{h}$ such that $\forall \xx\in L^*_{h},\,
\Tr^{-1}_{h}(c_{h})(\xx)=c_{0}(\xx-\yy_h)$. Denote by
$\Phi_0:A^{L^*_{0}}\to\az$ the bijective mapping putting in
correspondence any $c_0:L^*_{0}\to A$ with the configuration
$\Phi_{0}(c_0)\in\az$,
\[
 \left( c_{0}: L^*_{0}\to A \right)\quad \xrightarrow{\Phi_0}
 \quad \left(\Phi_{0}(c_{0}): \zdu\to A\right)
\]
such that $\forall t\in \z, \Phi_{0}(c_{0})(t):=
c_{0}(\varphi^{-1}(t))$.
The map $\Phi_0^{-1}: \az\to A^{L^*_{0}} $ associates any
configuration $a\in\az$ with the configuration $\Phi^{-1}_0(a)\in
A^{L^*_0}$ in the following way: $ \forall \xx\in L_0^*,\,
\Phi^{-1}_0(a)(\xx)=a(\varphi(\xx))$. Consider now the bijective
map $\Psi:\azdu\to (\az)^{\z}$ defined as follows
\[
\forall c\in\azdu, \quad \Psi(c)= (\ldots,
\Phi_0(\Tr_{-1}(c_{-1})), \Phi_0(\Tr_{0}(c_{0})),
\Phi_0(\Tr_{1}(c_{1})), \ldots)\enspace.
\]
Its inverse map $\Psi^{-1}:(\az)^{\z}\mapsto\azdu$ is such that
$\forall a\in(\az)^{\z}$,
\[
 \prec\Psi^{-1}(a)\succ= (\ldots,
\Tr_{-1}^{-1}(\Phi_0^{-1}(a_{-1})),
\Tr_{0}^{-1}(\Phi_0^{-1}(a_{0})),
\Tr_{1}^{-1}(\Phi_0^{-1}(a_{1})), \ldots)\enspace.
\]
Starting from a configuration $c$, the isomorphism $\Psi$ allows
to obtain a 1D configuration $a$ in which all components take
value from the same alphabet (see Figure~\ref{fig:same-alph}).

\begin{figure}[!htb]
  \begin{center}
     \includegraphics[scale=.4]{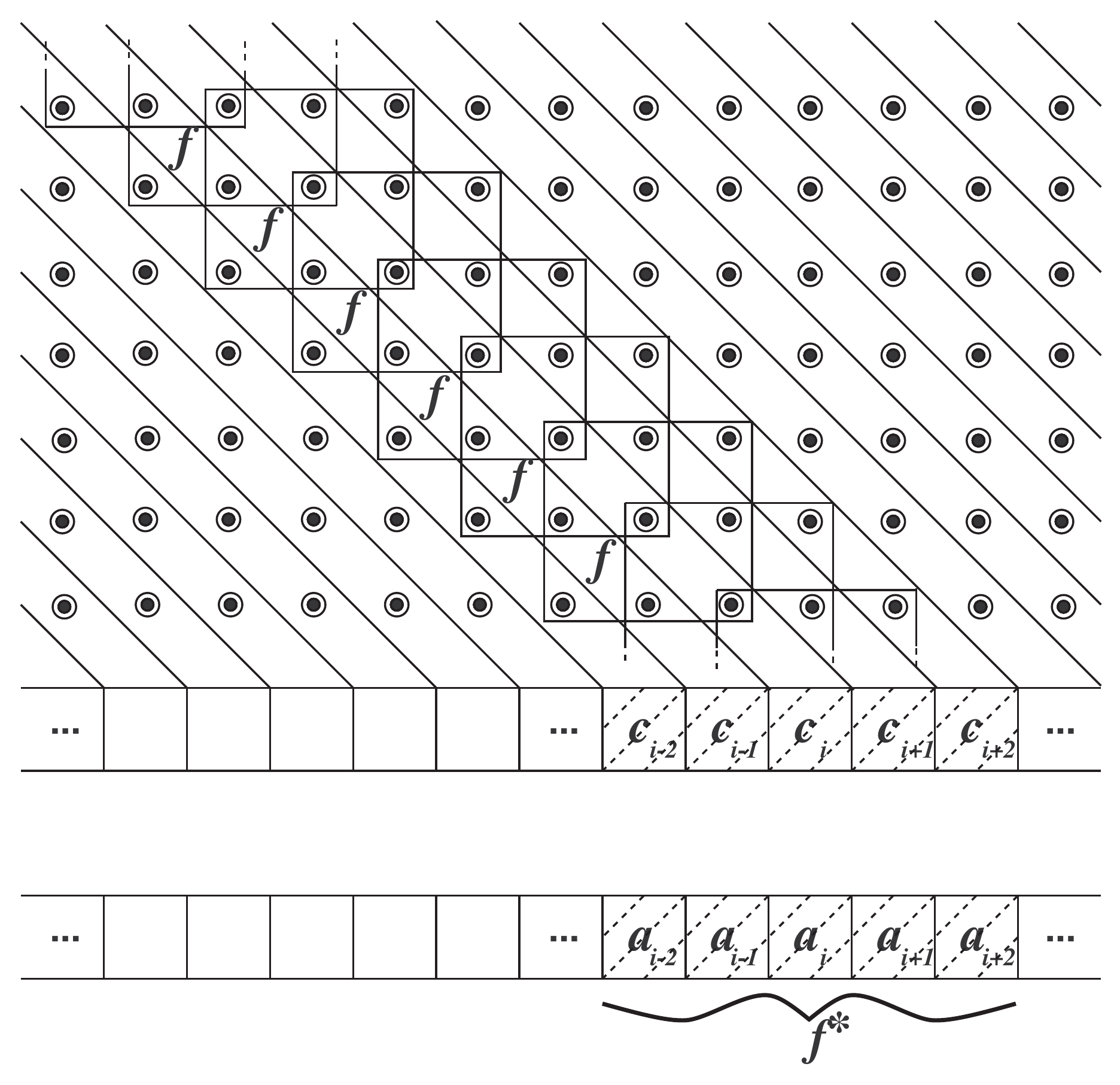}
   \end{center}
   \caption{Local rule $f^*$ of the 1D CA as sliced version of the original 2D CA.
   Here $r=1$ and $r^*=2$.}
   \label{fig:slicing-f}
\end{figure}

At this point, we have all the necessary formalism to correctly
define the radius $r^*$ local rule $f^*:(\az)^{2r^*+1}\to \az$
starting from a radius $r$ 2D CA $F$. Let $r_1$ and $r_2$ be the
indexes of the lines passing for $(r,r)$ and $(r,-r)$,
respectively. The radius of the 1D CA is $r^*=\max\set{r_1,r_2}$.
In other words, $r^*$ is such that $L_{-r^*},\ldots L_{r^*}$ are
all the lines which intersect the 2D $r$-radius Moore
neighborhood. The local rule is defined as
\[
\forall (a_{-r^*}, \ldots, a_{r^*})\in (\az)^{2r^*+1},\quad
 f^*(a_{-r^*}, \ldots , a_{r^*})= \Phi_0(b)
\]
where $b:L^*_0\to A$ is the slice obtained the simultaneous
application of the local rule $f$ of the original CA on the slices
$c_{-r^*},\ldots,c_{r^*}$ of any configuration $c$ such that
$\forall i\in[-r^*,r^*], c_i=\Tr^{-1}_i(\Phi^{-1}_0(a_i))$ (see
Figure~\ref{fig:slicing-f}). The global map of this new CA is
$F^*:(\az)^{\z}\to(\az)^{\z}$ and the link between $F^*$ and $f^*$
is given, as usual, by
\[
(F^*(a))_i=f^*(a_{i-r^*}, \ldots, a_{i+r^*})
\]
where $a=(\ldots, a_{-1}, a_0, a_1, \ldots)\in(\az)^{\z}$ and
$i\in\z$.
\smallskip

The slicing construction can be summarized by the following
\begin{theorem}\label{lem:slicingiso}
 Let $\para{\azdu, F}$ be a 2D CA and let $\para{(\az)^{\z},F^*}$ be the 1D CA
 obtained by the $\nn$ slicing construction of it, where $\nn\in\zdu$
 is a fixed vector. The two CA are isomorphic by the bijective
  mapping $\Psi$. Moreover,
  the map $\Psi^{-1}$ is continuous and then $\para{\azdu, F}$
  is a factor of $\para{(\az)^{\z},F^*}$.
\[
\begin{CD}
   (\az)^{\z}@>F^*>>&(\az)^{\z}\\
   @V{\Psi^{-1}}VV&@VV{\Psi^{-1}}V\\
   \azdu@>>F>&\azdu
\end{CD}
\]
\end{theorem}
\begin{proof}
It is clear that $\Psi$ is bijective.
We show that $\Psi\circ F=F^*\circ \Psi$, i.e., that $\forall
i\in\z, \forall c\in\azdu,\, \Psi(F(c))_i=F^*(\Psi(c))_i$. We have
$\Psi(F(c))_i=\Phi_0(\T_i(F(c)_i))$
where the slice $F(c)_i$ is obtained by the simultaneous
application of $f$ on the slices $c_{i-r^*},\ldots,c_{i+r^*}$. On
the other hand $F^*(\Psi(c))_i$ is equal to
\begin{eqnarray*}
f^*(\Psi(c)_{i-r^*}, \ldots, \Psi(c)_{i+r^*}) =
f^*(\Phi_0(\T_{i-r^*}(c_{i-r^*})),\ldots,
\Phi_0(\T_{i+r^*}(c_{i+r^*}))) =\Phi_0(b)
\end{eqnarray*}
where, by definition of $f^*$, $b$ is the slice obtained by the
simultaneous application of $f$ on the slices
\[
d_{r^*}=\T^{-1}_{-r^*}(\T_{i-r^*}(c_{i-r^*})), \ldots,
d_{r^*}=\T^{-1}_{r^*}(\T_{i+r^*}(c_{i+r^*}))
\]
which gives $b=\T_i(F(c)_i)$. We now prove that $\Psi^{-1}$ is a
continuous map from the 1D CA configuration space $(\az)^{\z}$ to
the 2D CA configuration space $\azdu$, both equipped with the
corresponding metric, which for the sake of simplicity is denoted
by the same symbol $d$. Choose an arbitrary configuration
$a=(\ldots, a_{-1}, a_0, a_{1}, \ldots)\in(\az)^{\z}$ and a real
number $\epsilon>0$. Let $n$ be a positive integer such that
$\frac{1}{2^n}<\epsilon$. Consider the lines $L^*_i$ which
intersect the 2D $n$-radius Moore neighborhood and let $m$ be the
maximum of the indexes of such lines.
Setting $\delta=\frac{1}{2^{m}}$, for any configuration
$b\in(\az)^{\z}$ with $d(b,a)<\delta$, we have that $b_i=a_i$ for
each integer $i\in[-m,m]$. This fact implies that
$(\Psi^{-1}(b))_i=(\Psi^{-1}(a))_i$ for each integer $i\in[-m,m]$,
and then $(\Psi^{-1}(b))_i(\xx)=(\Psi^{-1}(a))_i(\xx)$, for each
for any $\xx\in L^*_i$. Equivalently, we have
$(\Psi^{-1}(a))(\xx)=(\Psi^{-1}(b))(\xx)$, for any
$\xx\in\bigcup_{i\in[-m,m]} L^*_i$, and in particular for any
$\xx$ such that $|\xx|\leq n$. Hence,
$d(\Psi^{-1}(b),\Psi^{-1}(a))<\epsilon$ and $\Psi^{-1}$ is
continuous.
\end{proof}
\begin{remark}
The above constructions do not depend neither on the norm nor on
the sense of the vector $\nn$. In other words, if $\nn$ is a
normalized vector, all $k\nn$--slicing ($k\in\z$) constructions of
a CA $F$ generate the same CA $F^*$.
\end{remark}
\subsection{$\nn$-Slicing with finite alphabet}
Fix a vector $\nn\in\zdu$. For any 2D CA $F$, we can build an
associated sliced version $F^*$ with finite alphabet by
considering the $\nn$-slicing construction of the 2D CA restricted
on the set $\sv$, where $\vv$ is any vector such that
$\vv\perp\nn$. This is possible since the set $\sv$ is
$F$-invariant and so $(\sv,F)$ is a DTDS. The obtained
construction leads to the following
\begin{theorem}\label{lem:bz1d}
Let $F$ be a 2D CA and $\nn\in\zdu$. For any vector $\vv\in\zdu$
with $\vv\perp\nn$, the DTDS $\para{\sv, F}$ is topologically
conjugated to the 1D CA $\para{B^{\z}, F^*}$ on the finite
alphabet $B=A^{|\vv|}$ obtained by the $\nn$--slicing construction
of $F$ restricted on $\sv$.
\[
\begin{CD}
   B^{\z}@>F^*>>&B^{\z}\\
   @V{\Psi^{-1}}VV&@VV{\Psi^{-1}}V\\
   \sv@>>F>&\sv
\end{CD}
\]
\end{theorem}
\begin{proof}
Fix a vector $\vv\perp\nn$. Consider the slicing construction on
$\sv$. According to it, any configuration $c\in\sv$ is identified
with the corresponding bi-infinite sequence of slices. Since
slices of configurations in $\sv$ are in one-to-one correspondence
with symbols of the alphabet $B$, the $\nn$--slicing construction
gives a 1D CA $F^*:B^{\z}\to B^{\z}$ such that, by
Theorem~\ref{lem:slicingiso}, $\para{\sv, F}$ is isomorphic to
$\para{B^{\z}, F^*}$ by the bijective map $\Psi:\sv\to B^{\z}$. By
Theorem~\ref{lem:slicingiso}, $\Psi^{-1}$ is continuous. Since
configurations of $\sv$ are periodic with respect to
$\sigma^{\vv}$, $\Psi$ is continuous too.
\end{proof}
\begin{figure}[!htb]
  \begin{center}
     \includegraphics[scale=.4]{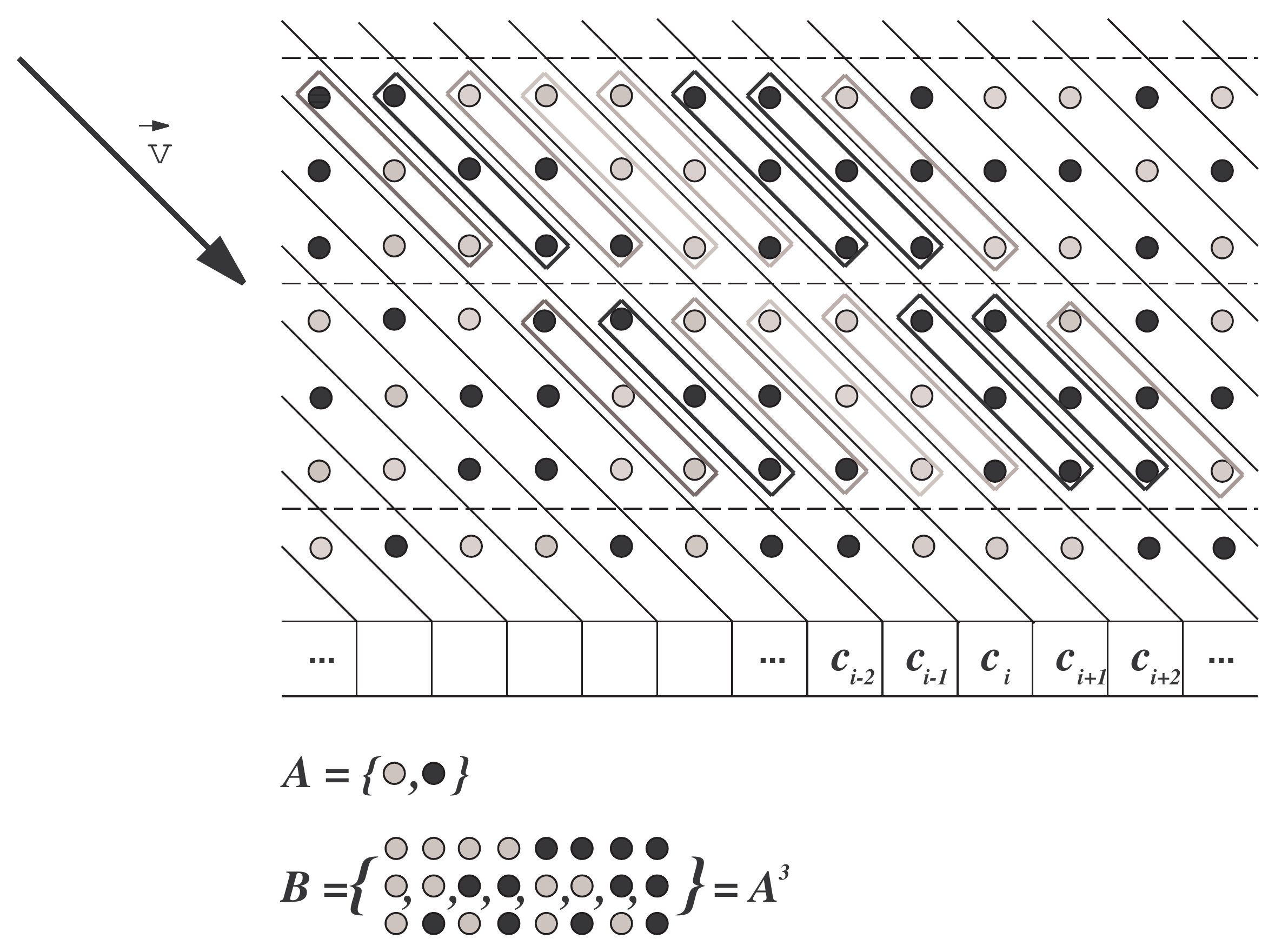}
   \end{center}
   \caption{$\nn$--slicing of a configuration $c\in\sv$
   on the binary alphabet $A$ where $\nn=(1,1)$ and
   $\vv=(3,-3)$. The configuration $\Psi(c)$ is on the alphabet $B=A^3$.}
   \label{fig:slifinito}
\end{figure}
The previous result is very useful since one can use all the
well-known results about 1D CA and try to lift them to $F$.
\section{Closingness and Openness for 2D CA.}\label{sec:clo-open}
The notion of closingness is of interest in 1D symbolic dynamics 
since it is tightly linked to several and important dynamical
behaviors. Moreover, it is a decidable property.  In this section, we generalize the definition of closingness to any direction
and we prove a strong relation w.r.t. openness.
\smallskip

\paragraph{Notation.} For any $\nn\in\zdu$, define $\bar{\nn}=-\nn$.
\begin{definition}[$\nn$-asymptotic configurations]
 Two configurations $c,c^\prime\in\azdu$ are
\emph{$\nn$-asymptotic} if there exists $q\in\z$ such that
$\forall \xx\in\zdu$ with $\nn\cdot\xx\geq q$
it holds that $c(\xx)=c'(\xx)$.
\end{definition}
\begin{definition}[$\nn$-closingness]
A 2D CA $F$ is $\nn$-closing if for any pair of
$\bar{\nn}$-asymptotic configurations $c,c^\prime\in\azdu$ we have
that $c\ne c'$ implies $F(c)\ne F(c')$. A 2D CA is \emph{closing}
if it is $\nn$-closing for some $\nn$.
\end{definition}
\ignore{
\begin{definition}[4-closingness]
A 2D CA $F$ is \emph{4-closing} if there exist a pair $(\ll, \mm)$
of independent vectors such that $F$ is $\nn$-closing for all
$\nn\in\set{\ll,\bar{\ll},\mm,\bar{\mm}}$).
\end{definition}
}
\ignore{
Remark that a 2D CA can be closing \wrt a certain direction but
may not be closing \wrt another one. For example, consider the 
radius $r=1$ 2D CA on the binary alphabet whose local rule 
performs the xor operator on the four corners of the Moore neighborhood. It is easy
to observe that this CA is $(1,1)$-closing but it is not
$(1,0)$-closing. \textbf{OCIO}}

\begin{definition}[$\nn$-$\mm$-closingness]
A 2D CA $F$ is \emph{$\nn$-$\mm$-closing} if for any pair of
$\bar{\nn}$-$\bar{\mm}$-asymptotic configurations (\ie
configurations which are both $\bar{\nn}$-asymptotic and 
$\bar{\mm}$-asymptotic)
$c,c'\in\azdu$, we have that $c\ne c'$ implies $F(c)\ne F(c')$.
\end{definition}

Thanks to the $\nn$-slicing construction with finite alphabet, the
following properties hold.
\begin{proposition}[\cite{dennunzio08,dennunzio09ja}]\label{lem:closing1d}
Let $F$ be a $\nn$-closing 2D CA. For any vector $\vv\in\zdu$ with
$\vv\perp\nn$, let $\para{B^{\z}, F^*}$ be the 1D CA of
Theorem~\ref{lem:bz1d} which is topologically conjugated to
$\para{\sv, F}$.
Then $F^*$ is either right or left closing.
\end{proposition}
\begin{theorem}[\cite{dennunzio08,dennunzio09ja}]\label{th:closingDPO}
  Any closing 2D CA has DPO.
\end{theorem}
Recall that a \emph{pattern} $P$ is a function from a finite domain
$Dom(P)\subseteq\zdu$ taking values in $A$. The notion of cylinder
can be conveniently extended to general patterns as follows: for
any pattern $P$, let $[P]$ be the set
\[
\set{c\in\azdu\;|\;\forall\xx\in Dom(P), c(\xx)=P(\xx)}\enspace.
\]
As in the 1D case, cylinders form a basis for the open sets.
For $h,t\geq 1$ and any normalized vectors $\nn,\mm\in\zdu$,
we say that a pattern $u$ has a $(\nn,\mm)$--\emph{shape} of size
$\sh{h,t}$ if for some $q,q^\prime\in\z$ it holds that
\[
dom(u)=\set{\xx\in\zdu\;|\;q\leq \nn\xx< q+h
\;\text{and}\; q^\prime\leq \mm\xx<q'+t}\enspace.
\] 
The following result is an improvement of~\cite[Thm. 2]{dennunzio08}
and gives a tight relation between closingness and
openess.
\begin{theorem}\label{th:2closingopen}
  If a 2D CA $F$ is both $\nn$ and $\bar{\nn}$--closing,
  then it is open.
\end{theorem}
\begin{proof}
We show that the image of any cylinder with $(\nn,\mm)$
shape is open, where $\mm\perp\nn$. Fix a cylinder $[u]$ where $u$
is a pattern centered in the origin and having a
$(\nn,\mm)$--shape of size $\sh{2h+1,2t+1}$. Let $\kk\perp \nn$
with $|\kk|=2t+1$ and denote $S^\prime_n=S_{n\kk}$. Consider the
dense set $S=\bigcup_{n\in\n} S^\prime_n$ endowed with the relative
topology ${\mathcal R}$. 
\smallskip

First of all, we prove that $F([u]\cap
S)$ is open in ${\mathcal R}$.
Choose a cylinder $[v]$ where $v$ is a pattern centered in
the origin and having a $(\nn,\mm)$--shape of size
$\sh{2l+1,2t+1}$ with $l>h$. In the sequel, we show that any
configuration from $[v]\cap S$ has a pre-image in $[u]$. If $c\in
[v]\cap S$ there exists $n$ such that $c\in S^\prime_n\cap
[v^\prime]$ where $[v^\prime]\subset [v]$ is the cylinder
individuated by a pattern $v^\prime$ having a $(\nn,\mm)$--shape
of size $\sh{2l+1,s}$ with $s=n|\vv|\geq 2t+1$. Let $\para{\bz,
F^*}$ be the 1D CA which is topologically conjugated to
$\para{S^\prime_n, F}$. By hypothesis and the slicing
construction, $\para{\bz, F^*}$ is both left and right closing.
Let $m>0$ be a integer from~\cite[Prop. 5.44]{kurka04}. Thus there
is a cylinder $[w]\subset [v^\prime]\subset [v]$ individuated by a
pattern $w$ having a $(\nn,\mm)$--shape of size $\sh{2m+1,s}$ and
such that $c\in [w]$. Equivalently, $\Psi(c)$ belongs to the 1D
cylinder $[\bar{w}]=\Psi([w]\cap S^\prime_n)\subset\bz$.
Using~\cite[Prop. 5.44]{kurka04} and a completeness argument, we
obtain that $\Psi(c)$ has a preimage in the 1D cylinder
$[\bar{u}]=\Psi([u]\cap S^\prime_n)\subset\bz$. This means that
$c$ has a preimage in $[u]$. Therefore, for a fixed integer $l>h$,
   \[F([u]\cap S)=\bigcup\set{[v]:
   F([u])\cap[v]\ne\emptyset,\;\text{and $v$ has $(\nn,\mm)$ shape of size}\;\sh{2l+1,2t+1}}
   \]
is a union of cylinders and hence $F([u]\cap S)$ is open in 
${\mathcal R}$.
\smallskip

It remains to prove that $F$ is open in the whole topology on
$\azdu$. Let $[u]$ be a cylinder and $c\in F([u])$. Since $S$ is
dense in $F([u])$, for any $r>0$ the ball $\mathcal{B}_r(c)$ of
center $c$ and radius $r$ contains a configuration $c^\prime\in
S$. In particular $\mathcal{B}_r(c)=\mathcal{B}_r(c^\prime)$.
Since $F$ is open in the relative topology $\mathcal{R}$, there
exists $\mathcal{B}_r(c^\prime)\cap S\subset F[u]\cap S$. Let
$b\in \mathcal{B}_r(c^\prime)$. Since $S$ is dense, there is a
sequence $\{b^{(n)}\}\in \mathcal{B}_r(c^\prime)\cap F([u])\cap S$
converging to $b$. Since $F([u])$ is closed, then $b\in F([u])$.
Thus, $\mathcal{B}_r(c^\prime)\subset F([u])$. \qed
\end{proof}
\begin{proposition}[\cite{dennunzio08,dennunzio09ja}]
   Any open 2D CA is surjective.
\end{proposition}

\section{Quasi-expansivity}
Shereshevsky proved that there are no positively expansive 2D CA~\cite{shereshevsky93}. Nevertheless, when watching the evolution of some 2D CA on a computer display, one can see many similarities with positively expansive 1D CA. Given two configurations, call \emph{defect}
any difference between them. Intuitively, a positively expansive CA is able to produce new defects at each evolution step and spread them to any direction of the cellular space. If in the 1D case, this is possible since there are only two directions (left and right), this is not
the case for CA over a 2D lattice where the number of possible directions is infinite.  In this section we introduce the notion of quasi-expansivity and we show that it shares with positive expansivity many
of the features just discussed.
\begin{definition}[Quasi--Expansivity]
\label{def:explike} A 2D CA $F$ is \emph{$\nn$--expansive} if the
1D CA $\para{(\az)^{\z},F^*}$ obtained by the $\nn$-slicing of it
is positively expansive. A 2D CA $F$ is \emph{quasi--expansive} if
it is $\nn$-expansive for some $\nn\in\zdu$.
\end{definition}
The following result follows from definition~\ref{def:explike} and
it will be useful in the sequel.
\begin{lemma}
\label{lem:nuexpansivef} Let $F$ be a $\nn$--expansive 2D CA. For
any vector $\vv\in\zdu$ with $\vv\perp\nn$, let $\para{B^{\z},
F^*}$ be the 1D CA of Theorem~\ref{lem:bz1d} which is
topologically conjugated to $\para{\sv, F}$.
Then $F^*$ is positively expansive.
\end{lemma}
\begin{theorem}
\label{th:expclosing} Any $\nn$--expansive 2D CA is both $\nn$ and
$\bar{\nn}$--closing.
\end{theorem}
\begin{proof}
Suppose that $F$ is not $\nn$--closing. Then, there exist two
distinct $\bar{\nn}$--a\-sym\-pto\-tic configurations $c,c^\prime$ such
that $F(c)=F(c^\prime)$. Let $\varepsilon$ be the expansivity
constant of the $\nn$--sliced CA $F^*$. By a shift argument, we
can assume that $d(\Psi(c),\Psi(c^\prime))<\varepsilon$. Thus, for
any $t\in\n$ it holds that $d(F^{*t}(\Psi(c)),
F^{*t}(\Psi(c^\prime)))<\varepsilon$. The proof for $\bar{\nn}$--closingness is similar.\qed
\ignore{
We show that $F$ is $\nn$--closing. Fix $\vv\in\zdu$ with
$\vv\perp\nn$ and for any integer $n>0$ denote
$S^\prime_n=S_{n\vv}$. By Lemma~\ref{lem:nuexpansivef}, for any
$n>0$, the CA $\para{\bz, F^*}$ conjugated to $\para{S^\prime_n,
F^*}$ is positively expansive and then, by a well known result
in~\cite{kurka04,nasu95}, it is both right and left closing. Thus, for
any $n>0$ and for any pair of distinct $\bar{\nn}$-asymptotic
configurations $a,b\in S^\prime_n$ it holds that $F(a)\neq F(b)$.
Let $c,c^\prime\in\azdu$ be two distinct $\bar{\nn}$-asymptotic
configurations. There exists a sequence $(a^{(n)},b^{(n)})\in
S^{\prime 2}_n$ of pairs of distinct $\bar{\nn}$-asymptotic
configurations converging to $(c,c^\prime)$ and such that
$F(a^{(n)})\neq F(b^{(n)})$. This assures that $F(c)\neq
F(c^\prime)$.
}
\end{proof}
\begin{corollary}\label{cor:quasiexp-dpo-surj-op}
Any quasi--expansive 2D CA has DPO, it is surjective and open.
\end{corollary}
\begin{proof}
It is an immediate consequence of Theorems~\ref{th:closingDPO},
\ref{th:2closingopen} and~\ref{th:expclosing}.\qed
\end{proof}
\begin{theorem}
\label{th:expmixing} Any quasi--expansive 2D CA $F$ is
topologically mixing.
\end{theorem}
\begin{proof}
Assume that $F$ is $\nn$--expansive. Choose $\varepsilon>0$ and
$c,c^{\prime}\in\azdu$. Take $\vv\in\zdu$ with $\vv\perp\nn$ and $e,
e^{\prime}\in\sv$ such that $d(c,e)<\varepsilon$ and
$d(c^{\prime}, e^{\prime})<\varepsilon$. Since $F$ is
$\nn$--expansive, by Lemma \ref{lem:nuexpansivef}, $\para{\sv, F}$
topologically conjugated to a 1D CA $\para{B^{\z},F^*}$ where
$F^*$ is positively expansive and $B$ is finite. Since positively
expansive 1D CA on a finite alphabet are topologically
mixing~\cite{kurka97,blanchard97}, there exist a sequence
$\{b^{(n)}\}\subset\sv$ and an integer $m\geq 0$ such that for all
$n\geq m$ it holds that $d(b^{(n-m)},e)<\varepsilon$ and
$d(F^n(b^{(n-m)}),e^\prime)<\varepsilon$. This concludes the
proof.\qed
\end{proof}
Let $\gv\in\{(1,1),(-1,1),(-1,-1),(1,-1)\}$.
We now give an example of a class of 2D CA which are quasi-expansive. 
\begin{definition}[Permutivity]
A 2D CA of local rule $f$ and radius $r$ is
\emph{$\gv$-permutive}, if for each pair of matrices
$N,N'\in\Mr$ with $N(\xx)=N'(\xx)$ in all vectors $\xx\neq
r\gv$, it holds that $N(r\gv)\ne N'(r\gv)$ implies
$f(N)\ne f(N')$. A 2D CA is \emph{bi-permutive} iff it is both
$\gv$ permutive and $\bar{\gv}$-permutive.
\end{definition}
The previous definition is given assuming a $r$ radius Moore
neighborhood. It is not difficult to generalize it to suitable
neighborhoods. The proofs of the results concerning permutivity
with different neighborhood can also be adapted.
\begin{proposition}[\cite{dennunzio08,dennunzio09ja}]\label{lem:permperm}
Consider a $\gv$-permutive 2D CA $F$. For any $\nn$ belonging either to the same quadrant or the opposite one as $\gv$, the 1D CA 
$\para{(\az)^\z, F^*}$ obtained by
the $\nn$-slicing construction is either rightmost or leftmost permutive.
\end{proposition}
\begin{lemma}
\label{lemma:bipinf} Let $\para{\az,F}$ be a 1D CA on a possibly
infinite alphabet $A$. If $F$ is both leftmost and rightmost
permutive, then $F$ is positively expansive.
\end{lemma}
\begin{proof}
We show that $F$ is positively expansive with constant
$\varepsilon=2^{-r}$ where $r$ is the radius of the CA. Choose
$c,c'\in\az$ with $c\neq c'$ and assume that for all $t\in\n$,
$F^t(c)_{[-r,r]}=F^t(c')_{[-r,r]}$. Suppose that $c_i\neq c'_i$ with
$i>r$. Let $n=\lfloor i/r\rfloor$ and $q=i-nr\in [0,r)$. Since
$F^n$ is rightmost permutive and $F^n(c)_q=F^n(c')_q$ then
$c_i=c'_i$. The case $c_i\neq c'_i$ with $i<-r$ is similar.\qed
\end{proof}
\begin{proposition}
\label{prop:bipexp} A 2D CA $F$ which is both $\gv$ and
$\bar{\gv}$--permutive is $\nn$-expansive for any $\nn$ belonging either to the same quadrant or the opposite one as $\gv$.
\end{proposition}
\begin{proof}
By Proposition~\ref{lem:permperm}, for any $\nn$ like in the hypothesis, the 1D CA $\para{(\az)^\z, F^*}$ obtained by
the $\nn$-slicing construction is both rightmost and leftmost
permutive. By Lemma~\ref{lemma:bipinf}, $\para{(\az)^\z, F^*}$ is
positively expansive and then $F$ is $\nn$--expansive.\qed
\end{proof}
Remark that 
a 2D CA can be $\nn$-expansive for a certain direction $\nn$ but not
for other directions as illustrated by the following example.
\begin{example}\label{ex:1}
Consider the 2D CA $F$ of radius $1$ on the binary alphabet
which local rule performs the xor operation on the two corners
$\gv=(1,1)$ and $\bar{\gv}=(-1,-1)$ of the Moore neighborhood.
Since $F$ is both $\gv$ and $\bar{\gv}$--permutive, $F$ is
$\nn$--expansive, and then $\nn$--closing, for all $ \nn$belonging either to the same quadrant or the opposite one as $\gv$. 
On the other hand, for $\nn=(1,-1)$ or
$\nn=(-1,1)$, $F$ is not $\nn$--closing and then not
$\nn$-expansive.\qed
\end{example}
\begin{proposition}
Any bipermutive 2D CA $F$ is open.
\end{proposition}
\begin{proof}
If $F$ is both $\gv$ and $\bar{\gv}$--permutive then, 
by \cite[Prop. 5]{dennunzio08}, it is both $\gv$ and $\bar{\gv}$--closing.
Theorem~\ref{th:2closingopen} concludes the proof.\qed
\ignore{
Assume that $F$ is both $\gv$ and $\bar{\gv}$--permutive.
Let $\mm,\ll\in\zdu$ be such that $\gv\cdot\mm>0$ and
$\gv\cdot\ll>0$. By Proposition~\ref{prop:bipexp}, $F$ is both
$\ll$ and $\mm$--expansive, and then, by
Theorem~\ref{th:expclosing}, it is $\ll,\bar{\ll},\mm,\bar{\mm}$ closing.
Theorem~\ref{th:4closingopen} concludes the proof.
}
\end{proof}

\subsection{Topological entropy of quasi-expansive CA}
The topological entropy is generally accepted as a measure of the complexity of a DTDS. The problem of computing (or even approximating) it for CA is algorithmically 
unsolvable~\cite{hurd92}. However, in~\cite{damico03}, the authors
provided a closed formula for computing the entropy of two important classes, namely additive CA and positively expansive CA.
In particular, they proved that for the first class, the entropy is either 
$0$ or $\infty$. Furthermore, in~\cite{mey08}, multidimensional cellular automata with finite nonzero entropy are exhibited. In this section, we shall see
another example of important class of CA with infinite topological entropy.

\paragraph{\emph{Notation.}} Given a 1D CA F and
$w,t\in\n$, let $N_F(w,t)$ be the number of distinct rectangles of width $w$ and height $t$ occurring in all possible space-time diagrams  
of $F$. 
Similarly, if $F$ is a D-dimensional CA, $N_F(w^{(D)},t)$ is the
number of distinct $D+1$ dimensional hyper-rectangles of
height $t$ and basis $w^{(D)}$, where 
$w^{(D)}$ is the $D$-dimensional hypercube of sides $w$.

In the case of D-dimensional CA, the definition of topological entropy for DTDS simplifies as follows~\cite{hurd92,damico03}:
\[
H(\azd,F)=\lim_{w\to\infty}\lim_{t\to\infty}\frac{\log N_F(w^{(D)},t)}{t}
\enspace.
\]
For introductory matters about topological entropy see~\cite{kurka04}.

\begin{theorem}\label{th:qe-infinite-tope}
Any quasi-expansive 2D (or higher) CA has infinite topological
entropy.
\end{theorem}
\proof
Consider a $\nn$--expansive 2D CA (for higher dimensions the proof is similar).
Fix a vector $\vv\in\zdu$ with $\vv\perp\nn$. For any $n\in\n$, let $S^\prime_n=S_{2^n\cdot\vv}$. 
By Lemma~\ref{lem:nuexpansivef} and Theorem~\ref{lem:bz1d}, any DTDS $\para{S^\prime_n, F}$ is topologically conjugated to a positively expansive 1D CA on a finite alphabet. 
By~\cite[Thm. 3.12]{nasu95}, each $\para{S^\prime_n, F}$ is also topologically conjugated to the DTDS $\para{(C_n)^\n,\sigma}$ for a suitable finite alphabet $C_n$. Thus, for any $n\in\n$, $H(S^\prime_n,F)=\log |C_n|$ where $|C_n|$ also represents the number of preimages of any element of $S^\prime_n$.
Since $S^\prime_n\subset S^\prime_{n+1}$, it holds that $H(S^\prime_n,F)\leq H(S^\prime_{n+1}, F)$. We show that $H(S^\prime_n,F)< H(S^\prime_{n+1}, F)$ for any $n\in\n$. This permits to conclude the proof since $H(S^\prime_n,F)\leq H(\azdu,F)$ for all $n\in\n$. 

For the sake of argument, assume that $H(S^\prime_n,F)= H(S^\prime_{n+1}, F)$ for some $n\in\n$. Thus, any element in $S^\prime_n$ has exactly $k$ pre--images in $S^\prime_n$ and any element in $S^\prime_{n+1}$ has exactly $k$ pre--images in $S^\prime_{n+1}$, where $k=|C_n|=|C_{n+1}|$. As a consequence, it holds that $F(S^\prime_{n+1}\setminus S^\prime_{n}) \subseteq S^\prime_{n+1}\setminus S^\prime_{n}$.  Since $\para{S^\prime_{n+1},F}$ is topologically conjugated to $\para{(C_{n+1})^\n,\sigma}$, it is also strongly transitive. 
Thus, if $c$ is any configuration in $S^\prime_n$ and $[u]$ is any 1D cylinder in $S^\prime_{n+1}\setminus S^\prime_{n}$, then $F(d)=c$ for some $d\in S^\prime_{n+1}\setminus S^\prime_{n}$ and $t\in\n$. Therefore $F(S^\prime_{n+1}\setminus S^\prime_{n})\cap S^\prime_{n}\neq \emptyset$ and this is a contradiction. 
\qed

\section{Quasi-almost equicontinuity vs. quasi-sensitivity}
In a similar way as quasi-expansivity, one can define quasi-sensitivity
and quasi-almost-equi\-conti\-nui\-ty.
\begin{definition}[Quasi-almost equicontinuity]
\label{def:almostlike} A 2D CA $F$ is \emph{$\nn$--almost
e\-qui\-con\-ti\-nuos} if the 1D CA $\para{(\az)^{\z},F^*}$ obtained by
the $\nn$ slicing of it is almost equicontinuous. A 2D CA $F$ is
\emph{quasi-almost equicontinuous} if it is $\nn$--almost
equicontinuous for some $\nn\in\zdu$.
\end{definition}
\begin{definition}[Quasi-sensitivity]
\label{def:senslike} A 2D CA $F$ is \emph{$\nn$--sensitive} if the
1D CA $\para{(\az)^{\z},F^*}$ obtained by the $\nn$ slicing of it
is sensitive. A 2D CA $F$ is \emph{quasi-sensitive} if it is
$\nn$--sensitive for some $\nn\in\zdu$.
\end{definition}
\begin{proposition}\label{prop:quasisens}
Any 2D CA $F$ is $\nn$-almost equicontinuous iff it is not
$\nn$-sensitive.
\end{proposition}
\proof
The ``only if'' part is obvious. For the opposite implication,
assume that $F$ is a non $\nn$--sensitive CA with radius $r$.
Then there exist $c\in (\az)^{\z}$ and $k\in\n$ such that for any
$c^\prime\in(\az)^{\z}$ with $c^\prime_{[-k,k]}=c_{[-k,k]}$ it
holds that $F^{*t}(c^\prime)_{[-r,r]}=F^{*t}(c)_{[-r,r]}$ for all
$t\in\n$. In particular, $u=c_{[-k,k]}$ is $r$--blocking for
$F^*$. For each $m\in\n$, define the open and dense set
$T(u,m)=\bigcup_{|i|\geq m} [u]_i\cap [u]_{-i}$. The set
$T(u)=\bigcap_{m\in\n} T(u,m)$ is also dense. We now show that any
$c\in T(u)$ is an equicontinuity point for $F^*$. Choose
$\varepsilon>0$ and let $j\in\n$ be such that
$\varepsilon<\frac{1}{2^j}$. There exist two integers $m\leq
-j-|u|$ and $n\geq j$ such that $c_{[m,m+|u|)}=c_{[n,n+|u|)}=u$.
Set $\delta=\min\{2^{m}, 2^{n+|u|}\}$ and take $c'\in(\az)^{\z}$
with $d(c^\prime,c)<\delta$. Since $u$ is $r$--blocking, for all
$t\in\n$ it holds that
$F^{*t}(c^\prime)_{[m+k-r,m+k+r]}=F^{*t}(c)_{[n+k-r,n+k+r]}$. This
fact assures that for each $m+k-r\leq i\leq n+k+r$ and any
$t\in\n$ $F^{*t}(c^\prime)_i=F^{*t}(c)_i$ and in particular
$d(F^{*t}(c^\prime), F^{*t}(c))<\varepsilon$.\qed

\begin{example}
Let $F$ and $\gv$ be as in Example~\ref{ex:1}. For any $\nn$ belonging
to the same quadrant or to the opposite one as $\gv$, $F$ is $\nn$--sensitive. However, for
$\ll=(-1,1)$ or $\ll=(1,-1)$, the CA $F^*$ can be seen as a CA of
radius $1$. Thus $F^*$ is equicontinuous and then $F$ is not
$\ll$--sensitive.\qed
\end{example}

\section{Closingness and undecidability}\label{sec:closund}

In this section we are going to prove the undecidability of $\mm$-closingness and $\mm$-$\nn$-closingness. These
results are obtained by adapting Kari's construction~\cite{kari94a}. First, we recall some basic definitions about tilings. Then, 
we briefly review Kari's construction to enlighten some details hidden in it which will be used in our proof. Afterwards, we modify it
to manage tilings stretched along non-orthogonal directions. Finally, the undecidability of closingness is proved.

\begin{figure}[t]
\begin{center}
\includegraphics[]{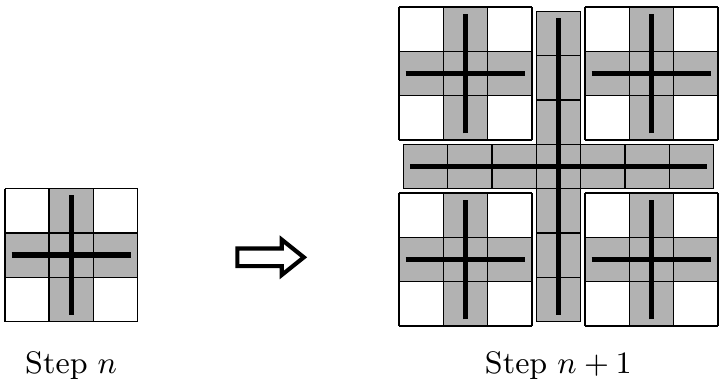}
\caption{The hierarchical structure used in Kari's construction. The largest (central) cross individuates the square.}
\label{fig:crocioni}
\end{center}
\end{figure}

\ignore{
\begin{figure}[t]
\begin{center}
\includegraphics[scale=0.32]{peano.pdf}
\caption{Example of plane-filling path.}
\label{ROB2}
\end{center}
\end{figure}
}

\paragraph{\emph{Tilings.}} We recall some basic notions about Wang 
tilings~\cite{wang62}.  A \emph{tile} is an oriented unit
square in which edges take a \emph{color} from a finite set $C$.  
A \emph{tile set} $\tau$ is a finite set of tiles with colors chosen from 
$C$. A tile set $\tau$ 
\emph{tiles the plane} if it is possible to arrange tiles from  $\tau$ over the grid $\z^2$ without rotations and in such a way that any two adjacent tiles respect the \emph{local color constraint} \ie they have 
the same color on  their common edge. A \emph{$\tau$-tiling}, or 
\emph{a tiling generated by $\tau$},  is a function from $\z^2$ to 
$\tau$ such that the local color constraints are respected.
A tile set is \emph{directed} if each tile is associated with a direction
in $\set{N,S,E,W}$. Tilings generated by directed tile sets define 
paths through the tiles in a natural way. The direction of
each tile tells which is the next tile visited in the path. A tiling generated by a directed tile set has the 
\emph{plane-filling property} if the path defined by it visits all the
tiles of arbitrary large squares.

In~\cite{B66}, Berger showed that Wang tilings can simulate
Turing machines in the sense that for any Turing machine $M$ and
any input $w$ there exists a tile set $\tau^{M,w}$ such that
$\tau^{M,w}$ tiles the plane if and only if $M$ does not halt on
input $w$. As a consequence, the problem
to establish whether a given tile set tiles the plane is undecidable. 

Remark that 2D CA can be seen as transformations on tilings.
Since most properties on tilings are undecidable, one might expect
that the same holds for properties on 2D CA. 
Indeed, Kari proved that this is the case for injectivity 
and surjectivity~\cite{kari94a}. We stress that these properties are decidable in dimension $1$.

\paragraph{\emph{Kari's construction.}} It is made of two parts
\begin{enumerate}
\item a tile set $k$ defining a hierarchical structure of ever-increasing squares of tiles;
\item directions are added to $k$ so that
there exists at least a $\tau$-tiling with the plane-filling property.
\end{enumerate}
\smallskip

Here we are not re-explaining Kari's construction in full details but just
give those details that are necessary in the sequel.
\smallskip

\noindent$(1)$\quad The hierarchical structure is defined recursively as follows.
For any $n\in\n$, the square of step $n+1$ consists in four copies
of squares of step $n$ separated by one horizontal and one vertical lines of suitable tiles which patterns form a big cross (see~\ref{fig:crocioni}).
Step $1$ consists in a square with a 3x3 central cross. All squares built up by this procedure respect local constraints. We omit details of
the specific tile set used, the interested reader can refer 
to~\cite{kari94a}. 

By compactness, this procedure grants that $k$ tiles the whole
plane $\z^2$. It is important to remark that (up to
translations) four different limit tilings can be obtained, depending on
the way the increasing squares are placed in the plane by
successive steps of the procedure.  

If at each step:
\begin{enumerate}
\item[i)] the SW corner of the new square is placed in the origin; then,
the obtained tiling contains only crosses with arms of finite length;
\item[ii)] the middle point of the south (resp., east) side of the new square is placed  in the origin; then, the obtained tiling contains a ``degenerated'' cross with a vertical (resp., horizontal) arm of infinite length and no horizontal (resp., vertical) arm;
\item[iii)] the new square is centered in the origin; then, the obtained tiling contains a cross with infinite vertical and horizontal arms. 
\end{enumerate}

Indeed, these were the very useful details hidden in Kari's proof.
In~\cite{kari94a}, only item i) is used.
\medskip

\par\noindent$(2)$\quad In the same way as Kari~\cite{kari94a}, we attach the classical Peano's curve \ignore{(Figure~\ref{ROB2})} to the hierarchical
structure defined in (1). We refer to~\cite{kari94a} for details on
how this can be done. Putting together (1) and (2), we may conclude
that for the case
\begin{enumerate}
\item[i)] the tiling contains a unique path visiting all tiles of $\z^2$; 
\item[ii)] the tiling contains two paths; each of them visits all tiles
of a half-plane;
\item[iii)] the tiling contains four paths; each of them visits all tiles of a quadrant. 
\end{enumerate}
 
\begin{figure}[t]
\begin{center}
\includegraphics[scale=.25]{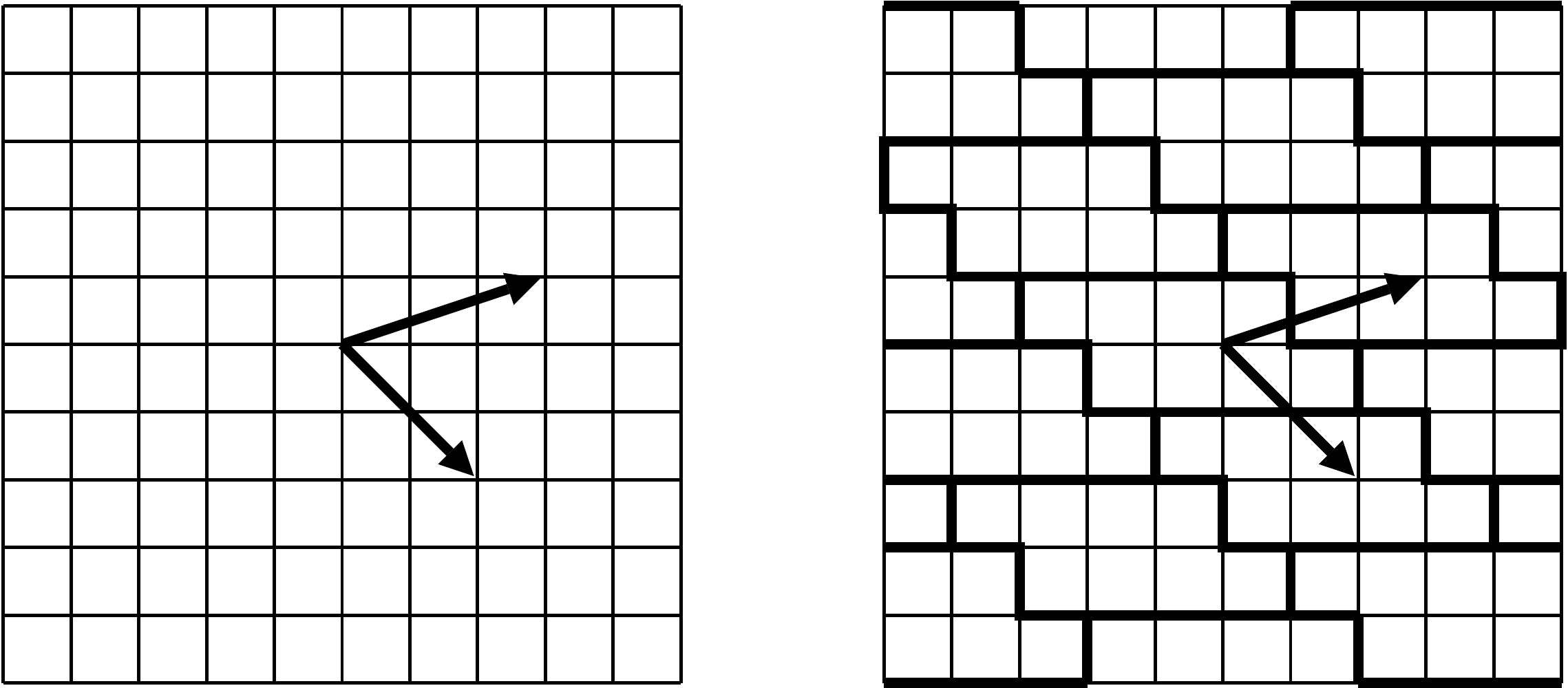}
\caption{Streching a tiling with respect to two directions.}
\label{fig:stretching-tiles}
\end{center}
\end{figure}


\paragraph{\emph{Stretching tiles.}}  We generalize the previous
construction in order to obtain paths visiting quadrants and halves-planes defined by any pair of directions.

Fix $\nn,\mm\in\z^2$. If tiles were not restricted to unit size squares and their shape could be changed, 
then it would be enough to transform
the tiles of the previous construction in parallelograms of sides $\nn$ and $\mm$ in order to reach our goal. This is not the case here,
therefore we should approximate parallelogram shapes
using a set of Wang tiles. 


Since $\nn$ and $\mm$ are integer vectors, there exists a 
connected shape  $\mathcal{N}$ such that $\mathcal{N}$ is made of tiles and it is possible to tile periodically the plane $\z^2$ by
patterns of domain $\mathcal{N}$
(see Figure~\ref{fig:stretching-tiles} as an example). 
The precise construction of $\mathcal{N}$ is easy but technical and
it is given in Appendix~\ref{app:tiling}. In the sequel, 
we call \emph{macro-tiles} each pattern of tiles of domain 
$\mathcal{N}$.

Macro-tiles have $4$ or $6$ neighboring macro-tiles, depending on
the angle between $\nn$ and $\mm$.
Given a macro-tile $T$, its North (resp., South) neighbor is the macro-tile pointed by $\nn$ (resp., $-\nn$); the East and West neighbors are defined similarly by $\mm$.
The remaining neighbors (if any) are called \emph{neutral} and
are denoted by $R_1,R_2$.
In particular, each macro-tile has $4$ sides (corresponding to North,
South, West or East neighbor) and possibly $2$ neutral sides (corresponding to neutral neighbors), see for example 
Figure~\ref{FIGAPP3}.
All definitions and properties of tilings extend in a natural way to 
tilings made by macro-tiles~\cite{LW08b,LW09b}.

We are going to color macro-tiles in
such a way that properties satisfied by $k$-tilings are also respected
by macro-tiles tilings. Let $n$ be the neutral color.

For any tile $t$ (resp., macro-tile $T$), $t(i)$ (resp., $T(i)$) is the
color of side $i$.
Given a tile $t\in k$, build the macro-tile $T_t$ of shape $\mathcal{N}$,
such that $T_t(i)=t(i)$ for $i\in\set{N,S,W,E}$; the remaining sides, if any are colored with 
$n$.  Moreover, $T_{t_1}(i)=T_{t_2}(i)=n$ for all $t_1,t_2\in k$ and
$i\in\set{R_1,R_2}$.
In other words, matching tiles in $k$ correspond to matching macro-tiles. 

\begin{figure}
\begin{center}
\includegraphics[scale=1.10]{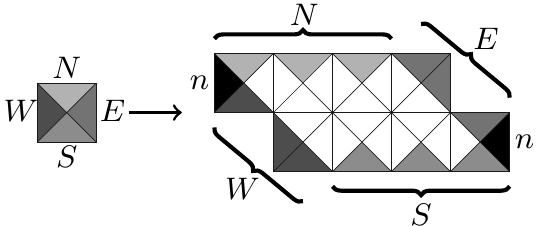}
\end{center}
\caption{Building macro-tiles from tiles.}
\label{fig:tiles-macro-tiles}
\end{figure}

Denote  $K_{\nn,\mm}$ the tile set which generates all the 
macro-tiles built in the above construction. See 
Figure~\ref{fig:tiles-macro-tiles} for a graphical illustration  
(macro-tiles are the same as in  Figure~\ref{fig:stretching-tiles}).
Since $k$ is a directed tile set, macro-tiles are also directed. 
Indeed, a $K_{\nn.\mm}$-tiling  defines a path that does not satisfy the plane-filling 
property but satisfies the following one:  the path visits all
patterns of domain $\mathcal{N}$ of arbitrary  large squares. 
We call this property the \emph{plane-pattern-filling property}.
We stress that the number of connected paths in $k$-tilings is same as the number
of plane-pattern-filling paths in $K_{\nn,\mm}$ tilings.

\paragraph{\emph{Back to closingness.}} We now have all the 
elements for proving the main result of this section.

\begin{theorem}\label{th:closingundec}
Let $\mm$ and $\nn$ be two vectors of $\Zgrid$. Then, $\mm$-closingness and $\nn$-$\mm$-closingness are undecidable for 2D CA.
\end{theorem}
\proof
For any tile set $\tau$, we build a 2D CA
$F_{\tau}$ such that the following equivalence holds: $F_{\tau}$ is $\mm$-closing
(resp., $\nn$-$\mm$-closing) if and only if $\tau$ does not tile the plane.

Cells of $F_{\tau}$ take a state in $K_{\nn,\mm}\times\tau\times B$, where
$B=\set{0,1}$ is the \emph{bit component} of the cell. Thus, a
configuration is the superposition of a $K_{\nn,\mm}$-tiling, a $\tau$-tiling (both possibly containing tiling errors) and a configuration in $\{0,1\}^{\Zgrid}$.

The CA $F_{\tau}$ has a Von Neumann
neighborhood of size $2\times m$, where $m$ is the size of the largest side of the macro-tile.
Therefore, the neighborhood of any tile of a macro-tile $T$ is big enough
to contain also the four neighboring macro-tiles of $T$.

The local rule $f$ does not change tiles but it possibly changes cell bit component. 
At each cell $\xx$ of $\Zgrid$, $f$ looks at the macro-tile containing $\xx$ and
its four neighboring macro-tiles. It verifies if both tilings are valid
 \ie if there is no two adjacent tiles with different colors on their common side.
 It also checks that all the cells in each of these five macro-tiles have the same bit component\footnote{In Kari's
construction, any tile has a bit. Here, since we are working with macro-tiles, we need to have the same bit component in all the cells of a macro-tile.}.
If both conditions are verified, $f$ changes the bit of $\xx$ 
by a $xor$ on it and the bit of cells in the macro-tile
pointed by the one containing $\xx$ (recall
that the macro-tile represents a tile of $k$ with a direction). Since all the bit
components of a macro-tile are the same, either they are all changed, or
none of them is changed. Otherwise, the bit of $\xx$ is left unchanged.

We now prove the equivalence. Assume that $\tau$ tiles the plane. Consider two  configurations $c$ and $c'$ as superpositions of the  
same valid $\tau$-tiling and the same $K_{\nn,\mm}$-tiling where the latter defines two (resp., four) plane-pattern-filling paths
separated by a line $d$ generated by $\nn$ (resp., lines $d$ and $d'$ generated by $\nn$ and
$\mm$). The bit components of $c$ and $c'$ are the same for any position
 on the right side of $d$
 (resp., on the right side of  $d$ and right side of $d'$). In all the other
positions they have value $0$ for $c$ and $1$ for $c'$. In this way, all the tiles of any macro-tile have the same bit component.  Moreover, $c$ and $c'$ are $\bar{\mm}$-asymptotic (resp.
$\bar{\nn}$-$\bar{\mm}$-asymptotic). 
Since both tilings are valid, 
the xor operates on all cells. The bits of $F_{\tau}(c)$ and $F_{\tau}(c')$ are the same for all cells on the right side of $d$ (resp. the quarter of plane delimited by $d$ and $d'$). Due to plane-pattern-filling paths, all bits of $F_{\tau}(c)$ and $F_{\tau}(c')$ have value $0$ in the other cells. Therefore, $F_{\tau}(c)=F_{\tau}(c')$ and $F_{\tau}$ is not
$\mm$-closing (resp. $\nn$-$\mm$-closing).

Conversely, if $F_{\tau}$ is not $\mm$-closing (resp. $\nn$-$\mm$-closing), there exist two different $\bar{\mm}$-asymptotic (resp. $\bar{\mm}$-$\bar{\nn}$-asymptotic)
configurations
$c$ and $c'$
such that $F_{\tau}(c)=F_{\tau}(c')$. The tiling components
of $c$ and $c'$ are the same since only the bits can be changed.
Let $\xx$ be a cell where the bits of $c$ and $c'$ are different.
Since $F_{\tau}(c)=F_{\tau}(c')$, both tiling components have to be valid in the
macro-tile containing $\xx$ and in its four neighboring macro-tiles. Moreover, 
the bit of $\xx$ has to be different from the bits of the macro-tile pointed by the
one containing $\xx$ (we are also sure that all the cells of both the macro-tiles
have the same bit since, in the opposite case, the bit in $\xx$ would not be changed). By repeating this argument on cells in the pointed macro-tile, we obtain that the tilings are valid in all macro-tiles of the plane-pattern-filling path.
 If the
$K_{\nn,\mm}$-tiling is valid along all macro-tiles of this infinite path, it means
that the $\tau$-tiling of $c$ and $c'$ is valid in arbitrary large squares (since a plane-pattern-filling
path visits all the macro-tiles of arbitrary large squares). 
Since $\tau$ tiles arbitrary big squares then, by compactness, it
tiles the plane. 

\section{Parallelograms and tiles}\label{app:tiling}

In this section we give full details on how to encode a parallelogram in tiles. Let $\nn$ and $\mm$ be two vectors of $\Zgrid$.
The goal is to build a pattern  $\mathcal{N}$ of shape more or less close to a parallelogram of
vectors $\nn$ and $\mm$, such that it is possible, with  $\mathcal{N}$, to tile the plane periodically with periods $\nn$
and $\mm$.

\smallskip

\begin{figure}
\begin{center}
\includegraphics{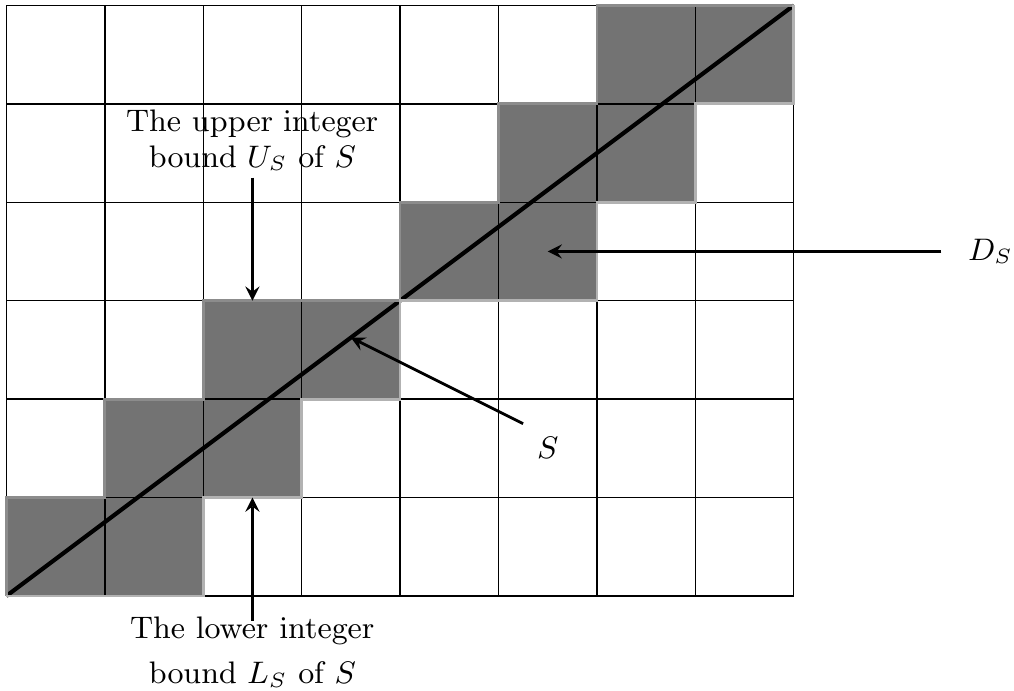}
\end{center}
\caption{A segment $S$, its integer approximation $D_S$ and the integer lower and upper bound.}
\label{FIGAPP1}
\end{figure}

\paragraph{Approximating a vector with tiles.} Let $\nn=(a,b)$ be a vector of $\Zgrid$.
We represent it by a segment $S$ going from $(0,0)$ to $(a,b)$. An integer unit size square of $\mathbb{R}^2$
is a square of size one with integer coordinates. We denote by $D_S$ the set of integer unit size squares of
$\mathbb{R}^2$ which have an intersection with $S$. The upper (resp. lower) integer bound $U_S$ (resp. $L_S$) of $S$ is the connected-path of $D_S$
coming from $(0,0)$ to $(a,b)$ which is an upper (resp. lower) bound of $D_S$. The Figure \ref{FIGAPP1} represents a segment $S$, its approximation $D_S$ and the two bounds.

Now, consider a parallelogram $\mathcal{A}$ (We stress that this transformation can be made for any
polygons with integer coordinate) of vectors $\nn$ and $\mm$. and denote by $a,b,c$ and $d$ its four sides. Without loss of generality we can assume that the vertexes of $\mathcal{A}$ are all at distance at least $3$ in both
vertical and horizontal directions. Indeed, if this is not the case, one can find the first integer $i$ such that the parallelogram of
vectors $i\times \nn$ and $i\times \mm$ has this property and make the same reasoning. The integer approximation $\mathcal{N}$ of the
parallelogram $\mathcal{A}$ is the polygon whose sides are the integer upper bounds $U_a, U_b, U_c$ and $U_d$ of the
sides $a,b,c$ and $d$.

We note that it is possible
that at a corner, if the angle between two sides of $\mathcal{A}$ is too little, that an overlapping
between two sides of its approximation $\mathcal{N}$ appears. In this case, we just suppress the overlapping
part to preserve the path-connected property. The suppression of this part does not affect the properties
of $\mathcal{N}$. The only difference is the following: in a periodic tiling with this shape, any pattern
have $6$ neighboring patterns rather that four in the case of no overlapping sides.

\begin{figure}
\begin{center}
\includegraphics{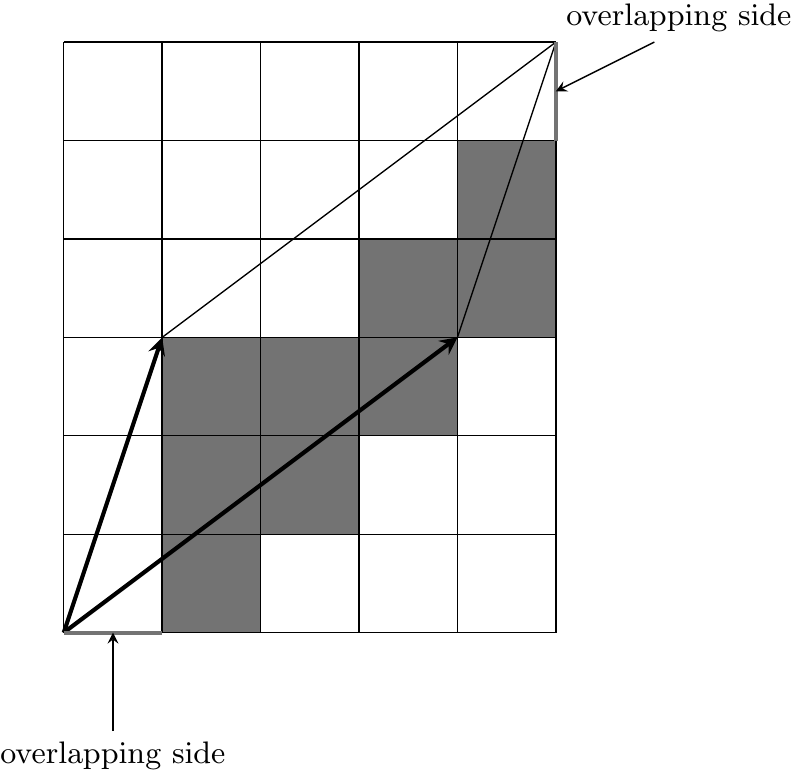}
\end{center}
\caption{An integer approximation of a parallelogram of vectors $(1,3)$ and $(4,3)$.}
\label{FIGAPP2}
\end{figure}

\begin{figure}
\begin{center}
\includegraphics{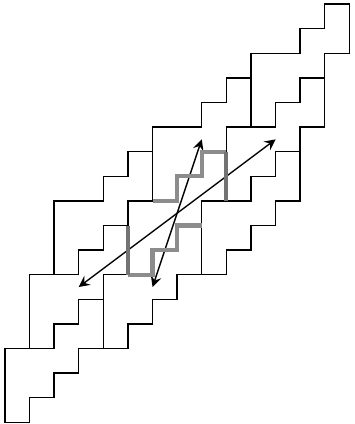}
\end{center}
\caption{The pattern $\mathcal{N}$ and its four translations by $\mm,\nn,-\mm$ and $-\nn$. In light grey and dark grey the common side with the neighbors.}
\label{FIGAPP3}
\end{figure}

 Since the lower and the
upper bound are the same for opposite sides, then two copies of the integer parallelogram $\mathcal{N}$ can be
assembled either on their east/west sides or on their north/south sides. Therefore, the shape $\mathcal{N}$ can
tile the plane periodically with period $\nn$ and $\mm$ (or multiple of these vectors).
The Figure \ref{FIGAPP2} shows a parallelogram of vectors $(1,3)$ and $(4,3)$, and its
integer approximation. This pattern contains two overlapping sides which are canceled. The Figure \ref{FIGAPP3}
shows that this pattern tiles the plane periodically with periods $(1,3)$ and $(4,3)$.

To stretch a tile set with respect to two directions $\nn$ and $\mm$, we use the integer approximation $\mathcal{N}$
of a parallelogram of vectors $\nn$ and $\mm$. The pattern $\mathcal{N}$ is path-connected, and can be tiled by
a tile set since it has only integer coordinates. We call {\em macro-tiles}, patterns of domain  $\mathcal{N}$.

Let $\tau$ be a tile set. Assume that $k$ tiles are needed to tile the pattern
$\mathcal{N}$. If $\tau$ contains $n$ tiles then its {\em stretched} version $\tau'$
is composed of $k.n$ tiles. Indeed, to each tiles $t=\{t(N),t(S),t(E),t(W)\}$ of $\tau$, we build $k$ tiles such that:

\begin{enumerate}[i)]
\item the $k$ tiles can be assembled only in an unique way to form a macro-tile of domain $\mathcal{N}$;
\item the colors of the sides of this macro-tile are the colors of the sides of the tile \ie the color
of the north side of the pattern \ie the common border with its north neighbor,
is $t(n)$ (the north color of $t$) and so on. If the pattern has overlapping sides, then there is some part of the border of the pattern
which is not in contact with one of its four neighbors. In this case, the color of these parts is neutral.

\end{enumerate}

The Figure \ref{FIGAPP3} shows the six neighbors of a pattern with overlapping sides. Four of them are its north,
south, east and west neighbors and their common borders are colored in gray and dark gray.

If two tiles of $\tau$ assemble on one side, then their corresponding $\tau'$-patterns assemble also on this
side: we have an isomorphism between the tiles of $\tau$ and the macro-tiles of $\tau'$ of domain $\mathcal{N}$.
One can see that for each $\tau$-tiling $P$, there exists a $\tau'$-tiling $P'$ which does the same as
$P$ but stretched with vectors $\nn$ and $\mm$.

\begin{figure}[t]
\begin{center}
\includegraphics{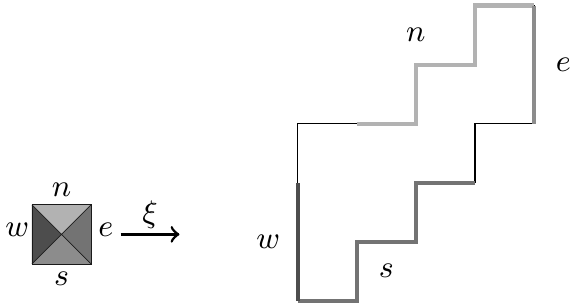}
\caption{The recursive transformation of a tile into a parallelogram.}
\label{FIGAPP4}
\end{center}
\end{figure}

The Figure \ref{FIGAPP4} illustrates the transformation of a tile in a pattern of domain $\mathcal{N}$. Only the border is
shown. The common side with the north pattern is colored with the north color of the tile and so on. The sides which
do not have a contact with one of the four neighboring macro-tiles are colored with a neutral color.

\end{document}